\documentclass[a4paper,12pt]{article}
\usepackage{graphicx,url,amsmath,amssymb}
\usepackage[nosectionbreak]{jcompbio}
\newcommand{\ignore}[1]{}

\title{Biological network comparison \\ via Ipsen-Mikhailov distance}
\DeclareRobustCommand{\authorthing}{
\small
\begin{center}
\begin{tabular}{ccc}
{\textsc{Giuseppe Jurman}${}^{*}$}&&{\textsc{Samantha Riccadonna}}\\
\multicolumn{3}{c}{Fondazione Bruno Kessler}\\
\multicolumn{3}{c}{via Sommarive 18 - Povo}\\
\multicolumn{3}{c}{I-38123 Trento (Italy)} \\
{\texttt{jurman@fbk.eu}} && {\texttt{riccadonna@fbk.eu}} \\
Ph:+39 0461314523 Fax: +39 0461314591&& Ph:+39 0461314650 Fax: +39 0461314591\\
\\
\multicolumn{3}{c}{\textsc{Roberto Visintainer}}\\ 
Fondazione Bruno Kessler && DISI, University of Trento \\
via Sommarive 18 - Povo && via Sommarive 14 - Povo\\
I-38123 Trento (Italy) && I-38123 Trento (Italy) \\
\multicolumn{3}{c}{\texttt{visintainer@fbk.eu}}\\
\multicolumn{3}{c}{Ph:+39 0461314650 Fax: +39 0461314591} \\
\\
\multicolumn{3}{c}{\textsc{Cesare Furlanello}}\\
\multicolumn{3}{c}{Fondazione Bruno Kessler}\\
\multicolumn{3}{c}{via Sommarive 18 - Povo}\\
\multicolumn{3}{c}{I-38123 Trento (Italy)} \\
\multicolumn{3}{c}{\texttt{furlan@fbk.eu}}\\
\multicolumn{3}{c}{Ph:+39 0461314580 Fax: +39 0461314591} \\
\end{tabular}
\vskip1cm
\end{center}
\normalsize
}
\preauthor{}
\author{\authorthing}
\postauthor{}
\date{}

\begin{document}
\maketitle

\section*{ABSTRACT}
\textbf{ 
Highlighting similarities and differences between networks is an informative task in investigating many biological processes. 
Typical examples are detecting differences between an inferred network and the corresponding gold standard, or evaluating changes in a dynamic network along time.
Although fruitful insights can be drawn by qualitative or feature-based methods, a distance must be used whenever a quantitative assessment is required.
Here we introduce the Ipsen-Mikhailov metric for biological network comparison, based on the difference of the distributions of the Laplacian eigenvalues of the compared graphs.
Being a spectral measure, its focus is on the general structure of the net so it can overcome the issues affecting local metrics such as the edit distances.
Relation with the classical Matthews Correlation Coefficient (MCC) is discussed, showing the finer discriminant resolution achieved by the Ipsen-Mikhailov metric.
We conclude with three examples of application in functional genomic tasks, including stability of network reconstruction as robustness to data subsampling, variability in dynamical networks and differences in networks associated to a classification task.
}

\textbf{Key words:} Network comparison, Network distance, Graph spectrum, Laplacian matrix.

\section{INTRODUCTION}
\label{sec:intro}
Networks methods in biology have recently gained popularity among researchers worldwide and they are nowadays pervading a relevant portion of scientific literature: see \citep{pavlopoulos11using} for a recent review and \citep{buchanan10networks} for a comprehensive reference.
Their role is believed to have an even higher impact in future: a good example is the case of network medicine \citep{barabasi11network,vidal11interactome}.
A central problem is the comparison of biological networks, a task occurring in many areas of biology.
Examples include detecting similarities in gene networks related to the same pathway across different species, or tracking the evolution of the network wiring during a biological process, or highlighting variations between networks associated to different pathophysiological conditions.
Classical comparison measures are the pairs Precision/Recall or Sensitivity/Specificity, or the F-score (for instance in network reconstruction), or the Maximal Common Subgraph distance (in network alignment).
More recently, the use of the Matthews Correlation Coefficient (MCC) \citep{maqc10maqcII} has been borrowed from the machine learning community as a more reliable indicator for summarizing the confusion matrix into a single figure.
Other cost-based functions stem from the theory of graph matching: the edit distance and its variants use the minimum cost of transformation of one graph into another by means of the usual edit operations - insertion and deletion of links.
These similarity measures are widely considered in literature: see \citep{bunke00graph} for an introductory review.
In alternative, the theory of network measurements relies on the quantitative description of main properties such as degree distribution and correlation, path lenghts, diameter, clustering, presence of motives. 
However, all these measures are local, because, for each link, only the structure of its neighbourhood gives a contribution to the distance value, while the structure of the whole topology is not considered.
To overcome the locality issue in network comparison, a few global distances have been proposed: among them, the family of spectral measures is particularly relevant. 
As the name suggests, their definition is based on (functions of) the spectrum of one of the possible connectivity matrices of the network, \textit{i.e.} its set of eigenvalues. 
The spectral theory has been applied to biological networks in \citep{banerjee09graph}, where the properties of being scale-free\footnote{Scale-freeness: the degree distribution follows a power law.} and small-world\footnote{Small-world nets: most nodes are not neighbors of one another, but most nodes can be reached from every other by a small number of hops or steps.} are particularly evident. 
Isospectral networks cannot be distinguished by this class of measures, so all these measures are indeed distances between classes of isospectral graphs: however, the number of isospectral networks is negligible for large number of nodes \citep{haemers04enumeration}. 
In a recent paper \citep{jurman11introduction}, we described six spectral distances, showing their behaviour on synthetic benchmarks and on the transcriptional network of \textit{E. coli} from the RegulonDB\footnote{\url{http://regulondb.ccg.unam.mx/}} database.
On the ground of such experiments, we choose Ipsen-Mikhailov $\epsilon$ distance \citep{ipsen02evolutionary} out of the six original metrics for stability and robustness. 
In \citep{barla11machine} we show a complete functional genomic pipeline employing Ipsen-Mikhailov metric for the detection of the discriminant pathways after a machine learning preprocessing.
The $\epsilon$ metric evaluates the difference of the distribution of Laplacian eigenvalues between two networks: as such, it can also be interpreted as a measure of the different network synchronizability \citep{belykh05synchronization,wu08synchronization}.
Here we show the relation of $\epsilon$ distance with MCC, and we present examples of application for network comparison in situations of biological interest such evolving dynamical network and comparison of miRNA networks associated to predictive discrimination in hepatocellular carcinoma.
Finally, we also show the use of Ipsen-Mikhailov distance in evaluating the stability of the reconstruction of a network from microarray data in terms of robustness to data subsampling, in order to quantitatively express the level of reliability of a given inference.

\section{IPSEN-MIKHAILOV $\pmb{\epsilon}$ DISTANCE}
\label{sec:ipsen}
Originally introduced in \citep{ipsen02evolutionary} as a tool for network reconstruction from its Laplacian spectrum, the definition of the Ipsen-Mikhailov $\epsilon$ metric follows the dynamical interpretation of a $N$-nodes network as a $N$-atoms molecules connected by identical elastic strings, where the pattern of connections is defined by the adjacency matrix of the corresponding network.
The dynamical system is described by the set of $N$ differential equations 
\begin{equation}
\label{eq:ipsen_model}
\ddot{x}_i+\sum_{j=1}^N A_{ij}(x_i-x_j)=0\quad\textrm{for\;$i=0,\cdots,N-1$}\ .
\end{equation}
We recall that the Laplacian matrix $L$ of an undirected network is defined as the difference between the degree $D$ and the adjacency $A$ matrices $L=D-A$, where $D$ is the diagonal matrix with vertex degrees as entries.
$L$ is positive semidefinite and singular \citep{chung97spectral,atay06network,spielman09spectral,tonjes09perturbation,atay06network}, so its eigenvalues are $0 = \lambda_0 \leq \lambda_1\leq \cdots\leq \lambda_{n-1}$.
The vibrational frequencies $\omega_i$ for the network model in Eq.~\ref{eq:ipsen_model} are given by the eigenvalues of the Laplacian matrix of the network: $\lambda_i = \omega^2_i$, with $\lambda_0=\omega_0=0$. 
In \citep{chung97spectral}, the Laplacian spectrum is called the vibrational spectrum.
Estimates (also asymptotic) of the eigenvalues distribution are available for complex networks \citep{rodgers05eigenvalue}.  
Moreover, the relation between the spectral properties and the structure and the dynamics of a network are discussed in \citep{jost02evolving, jost07dynamical, almendral07dynamical}.

The spectral density for a graph as the sum of Lorentz distributions is defined as 
\begin{displaymath}
\rho(\omega)=K\sum_{i=1}^{N-1} \frac{\gamma}{(\omega-\omega_k)^2+\gamma^2}\ ,
\end{displaymath}
where $\gamma$ is the common width and $K$ is the normalization constant solution of $\displaystyle{\int_0^\infty \rho(\omega)\textrm{d}\omega =1}$.
The scale parameter $\gamma$ specifies the half-width at half-maximum, which is equal to half the interquartile range. 
It works as a multiplicative factor for the distance and in all experiments hereafter, $\gamma$ is set to $0.08$ as in the original reference. 

Then the spectral distance $\epsilon$ between two graphs $G$ and $H$ with densities $\rho_G(\omega)$ and $\rho_H(\omega)$ can then be defined as 
\begin{displaymath}
\epsilon(G,H) = \sqrt{\int_0^\infty \left[\rho_G(\omega)-\rho_H(\omega)\right]^2} \textrm{d}\omega\ .
\end{displaymath}
Because of the definition of Ipsen-Mikhailov distance, a comparison can be computed only between networks with the same (number of) nodes.
In order to get rid of the intrinsic dependence of the distance of the number of nodes of the compared networks, a normalization factor can be introduced, defined as the distance between $E_n$ and $F_n$, respectively the totally disconnected and the fully connected graph on $n$ nodes:
\begin{displaymath}
\hat{\epsilon}(G,H) = \frac{\epsilon(G,H)}{\epsilon(E_n,F_n)}\ ,
\end{displaymath}
for $n$ the number of nodes of $G$ and $H$.

\section{RELATION WITH THE MATTHEWS CORRELATION COEFFICIENT}
\label{sec:mcc}
We first compare $\epsilon$ with Matthews Correlation Coefficient (MCC for short), a measure of common use in the machine learning community \citep{baldi00assessing} and recently accepted as an effective metric also for network comparison \citep{supper07reconstructing,stokic09fast}.
The MCC allows summarizing into a single value the confusion matrix of a binary classification task, thus working as a reliable alternative to measures obtained as functions of Sensitivity/Specificity and Precision/Recall.
Originally introduced in \citep{matthews75comparison}, it is also known as the $\phi$-coefficient, corresponding for a $2\times 2$ contingency table to the square root of the average $\chi^2$ statistic 
\begin{displaymath}
\textrm{MCC}=\sqrt{\chi^2 / N}\ ,
\end{displaymath}
where $N$ is the total number of observations.
As an example of use in bioinformatics, MCC  has been chosen as the reference metric in the US FDA-led initiative MAQC-II aimed at reaching consensus on the best practices for development and validation of predictive models based on microarray gene expression and genotyping data for personalized medicine \citep{maqc10maqcII}.
In the binary case of two classes positive ($P$) and negative ($N$), for the confusion matrix $\left(\begin{smallmatrix} \textrm{TP} & \textrm{FN} \\ \textrm{FP} & \textrm{TN}\end{smallmatrix}\right)$, where $T$ and $F$ stand for true and false respectively, the Matthews Correlation Coefficient has the following shape:
\begin{displaymath}
\textrm{MCC} = \frac{\textrm{TP}\cdot\textrm{TN}-\textrm{FP}\cdot\textrm{FN}}{\sqrt{\left(\textrm{TP}+\textrm{FP}\right)\left(\textrm{TP}+\textrm{FN}\right)\left(\textrm{TN}+\textrm{FP}\right)\left(\textrm{TN}+\textrm{FN}\right)}}\ .
\end{displaymath}
MCC lives in the range $[-1,1]$, where $1$ is perfect classification, $-1$ is reached in the complete misclassification case while $0$ corresponds to coin tossing classification.
Note that MCC is invariant for scalar multiplication of the whole confusion matrix.

We compare $\epsilon$ and MCC in two synthetic network experiments.

\begin{figure}[!ht]
\includegraphics[width=0.95\textwidth, angle=-90]{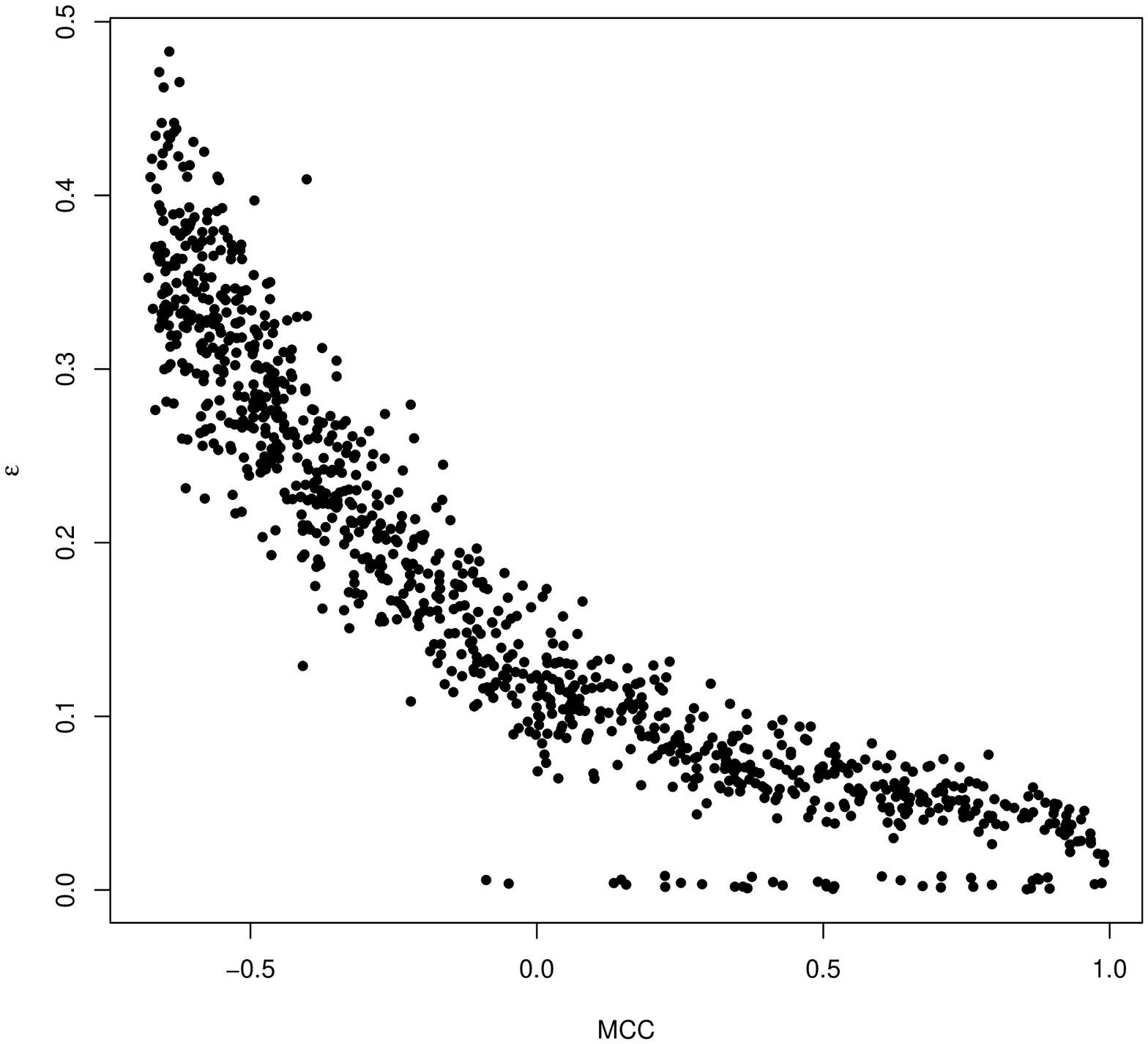}
\caption{MCC versus $\epsilon$ distance for 1000 pairs of randomly generated topologies on 1000 nodes.}
\label{fig:mcc_im}
\end{figure}

First we generate 1000 pairs of network topologies $(N_1,N_2)$ on $n = 1000$ nodes as follows. 
The adjacency matrix for $N_1$ is randomly generated by associating to each of the $\binom{n}{2}=\frac{n(n-1)}{2}=4950$ possible links a weight $w$ sampled by a uniform distribution in the unit interval: a link is then declared existing whenever $w>0.75$. 
The network $N_2$ is generated by rewiring $N_1$ through deletion of $p_1$\% of the existing links and insertion of $p_2$\% novel links, for $p_1$ and $p_2$ uniformly sampled in $[0,90]$. 
Then, for each pair $(N_1,N_2)$, we compute the MCC and the $\epsilon$ metrics: the results are displayed in Fig.~\ref{fig:mcc_im}. 
The plot suggests that, although there is a coherent trend between the two measures, the variability is quite high: (anti)correlation value for the two measures is 0.901. 

\begin{figure}[!ht]
\begin{center}
\begin{tabular}{ccc}
\includegraphics[width=0.3\textwidth]{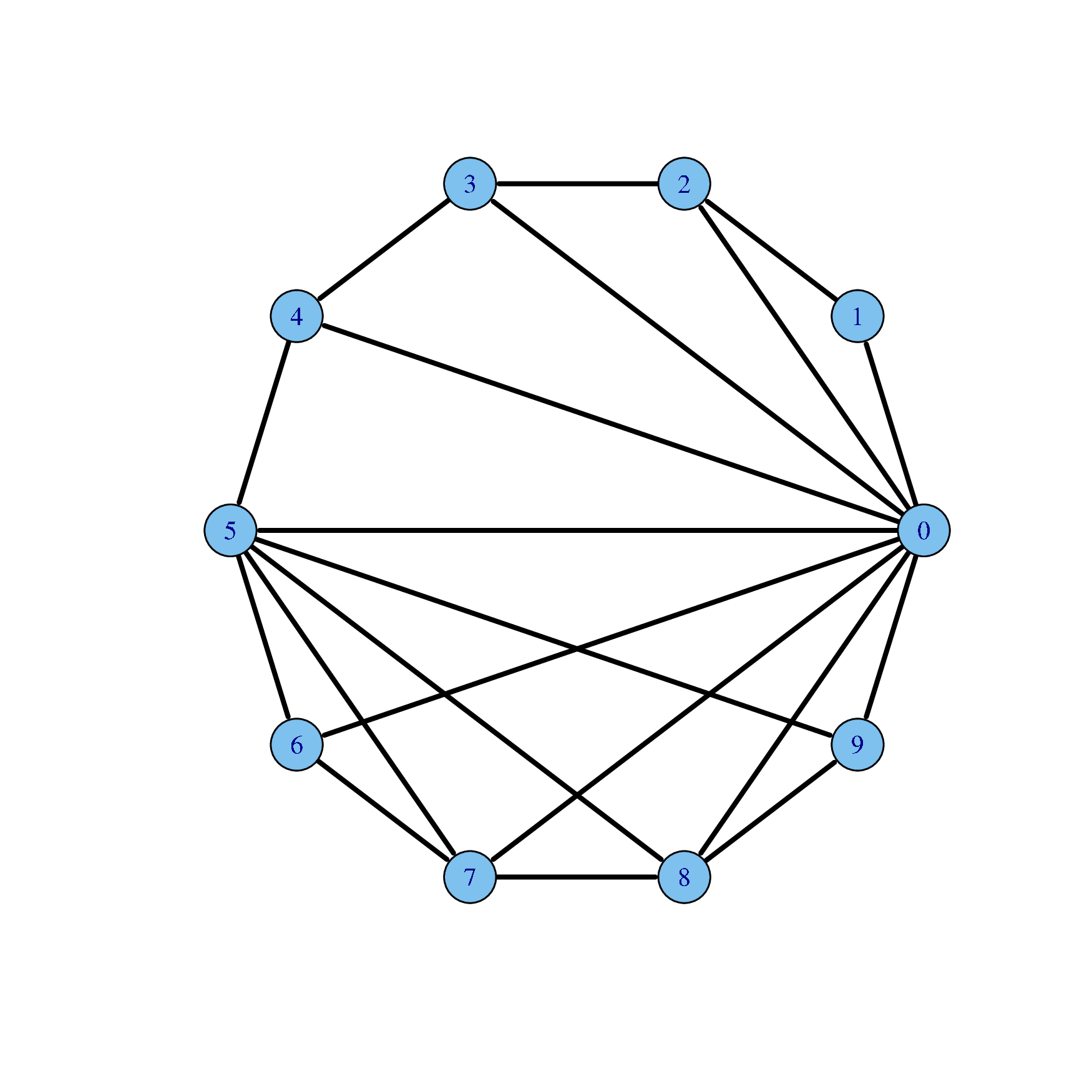}&
\includegraphics[width=0.3\textwidth]{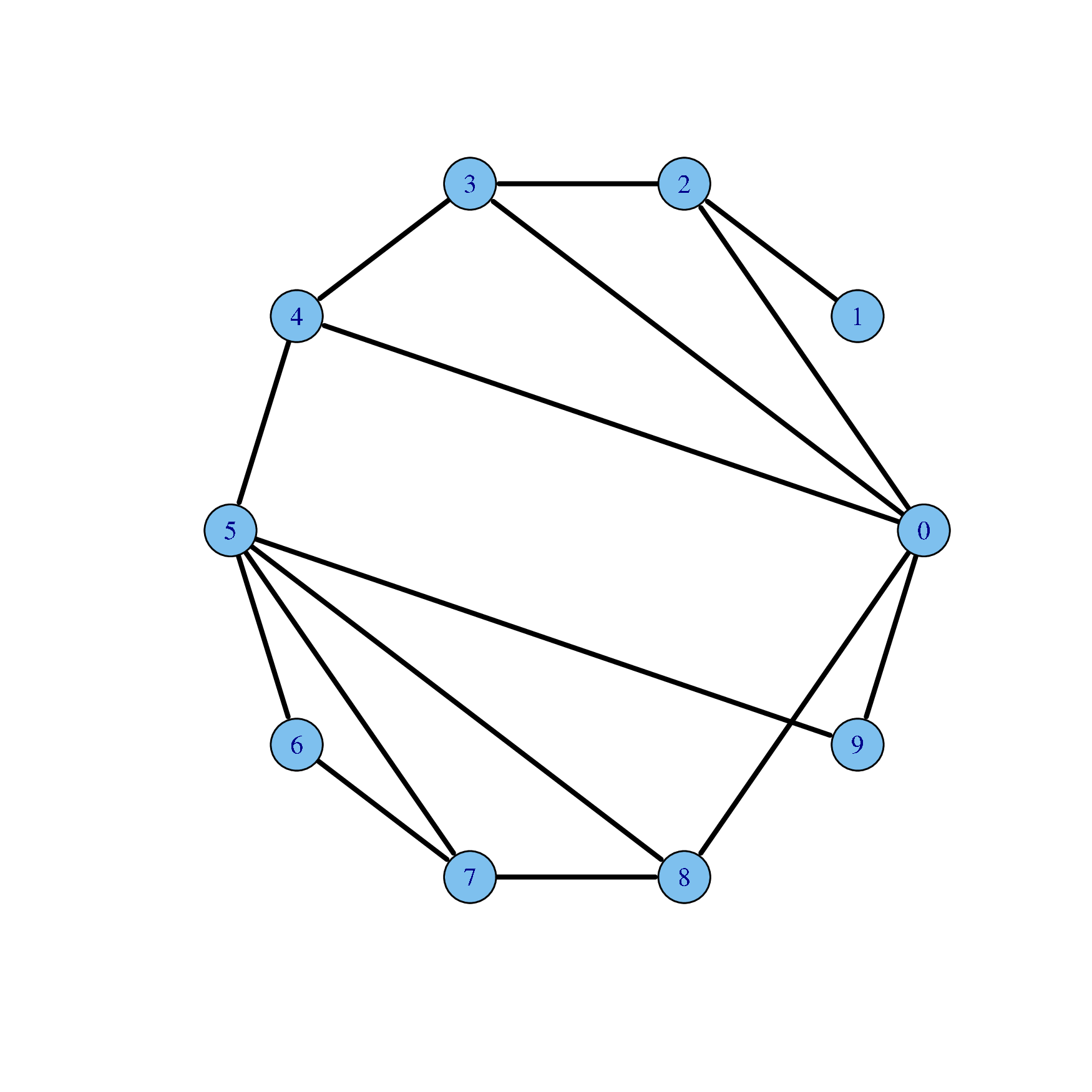}&
\includegraphics[width=0.3\textwidth]{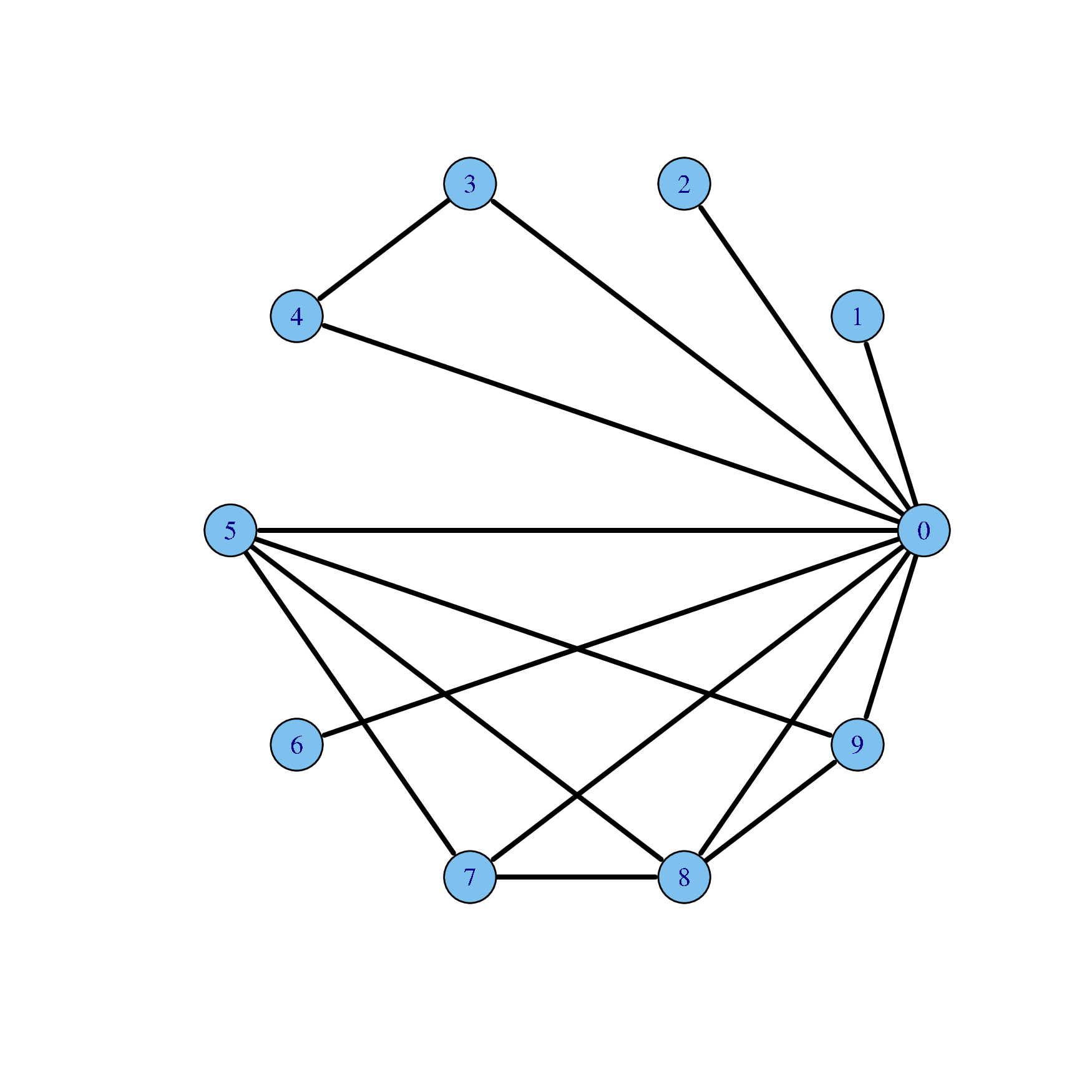}\\
(a) $N$ & (b) $N_\textrm{min}$ & (c) $N_\textrm{max}$
\end{tabular}
\end{center}
\caption{(a) The ground truth network, (b) the network $N_\textrm{min}$ with minimal Ipsen-Mikhailov distance and (c) the network $N_\textrm{max}$ with maximal distance from the ground truth.}
\label{fig:n_max_min}
\end{figure}

The second experiment is aimed at quantifying the detected variability.
A simple network $N$ is created on 10 nodes with 20 links (of 45 potential) to be used as the ground truth: its topology is displayed in Fig. \ref{fig:n_max_min}(a). 
Then a set of 1000 networks $\mathcal{S} = \{N_i\}_{i=1}^{1000}$ is created from the topology of $N$ by randomly deleting 5 links (the total number of all such networks is $\binom{20}{5}= 9302400$). 
All elements of $\mathcal{S}$ have confusion matrix 
$\left(\begin{smallmatrix} \textrm{TP} & \textrm{FN} \\ \textrm{FP} & \textrm{TN}\end{smallmatrix}\right) =
\left(\begin{smallmatrix} \textrm{15} & \textrm{0} \\ \textrm{5} & \textrm{25}\end{smallmatrix}\right) $
and thus for each $N_i\in\mathcal{S}$, 
$\textrm{MCC}=\frac{15\cdot 25}{\sqrt{20\cdot 15\cdot 3 0\cdot 25}}=\frac{\sqrt{10}}{4}\approx 0.79$.

\begin{figure}[!ht]
\includegraphics[width=0.95\textwidth]{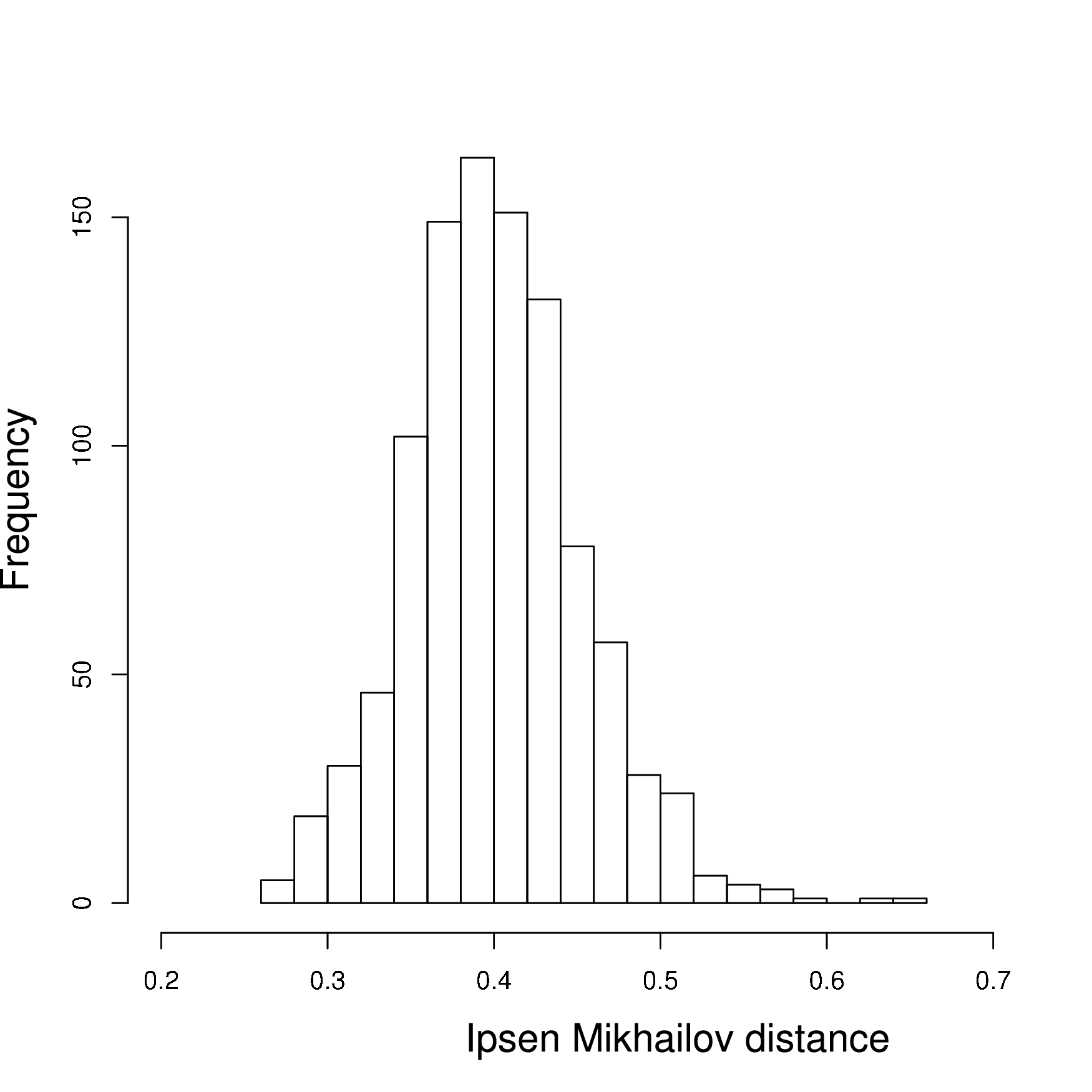}
\caption{Histogram of 1000 values of Ipsen-Mikhailov values for networks with fixed $MCC=\frac{\sqrt{10}}{4}$ distance from the given ground truth in Fig.~\ref{fig:n_max_min}(a).}
\label{fig:hist}
\end{figure}

For each $N_i$, the corresponding distance to the ground truth $\epsilon(N_i,N)$ is computed: the corresponding histogram of the 1000 values of the Ipsen-Mikhailov distance is shown in Fig.~\ref{fig:hist}.
As expected, the variability in the obtained values for $\epsilon$ is very high: the range is [0.2670, 0.6438], with mean 0.4010 and median 0.3977. 
This result shows that the topologies of networks with a given confusion matrix can be structurally very different. 
For instance, in Fig.~\ref{fig:n_max_min}(b,c) we show the two networks $N_\textrm{min}$, $N_\textrm{max}$ associated to extremal values of $\epsilon$.
Both these experiments support the claim that Ipsen-Mikhailov metric has an higher resolution in discriminating between network structures.

\section{APPLICATIONS}
\label{sec:results}
\subsection*{Evolution of dynamic networks}
\begin{figure}[!t]
\begin{center}
\begin{tabular}{cc}
\hskip-9cm\includegraphics[width=0.49\textwidth, angle=-90]{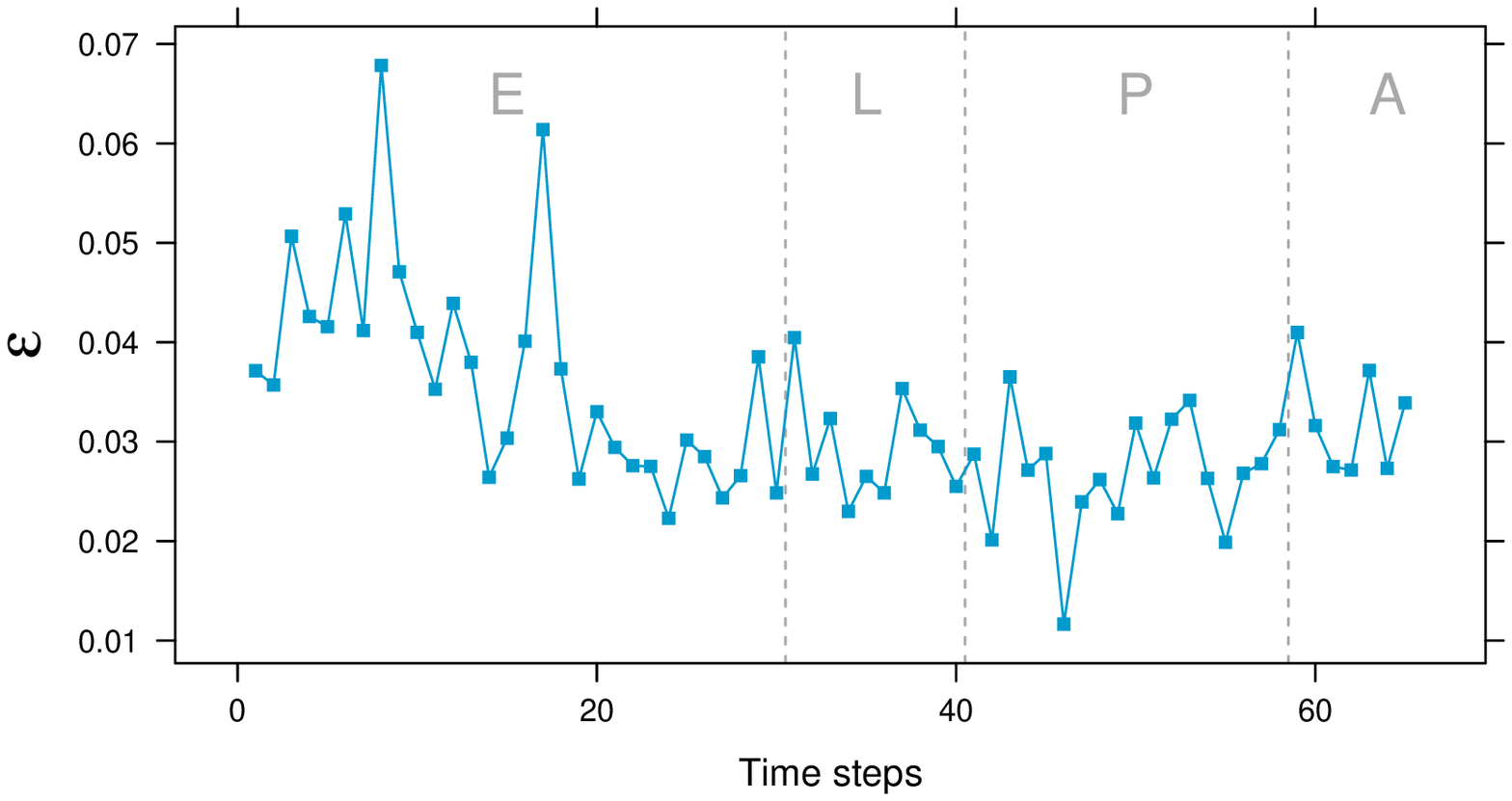} &\hskip-9cm\includegraphics[width=0.49\textwidth, angle=-90]{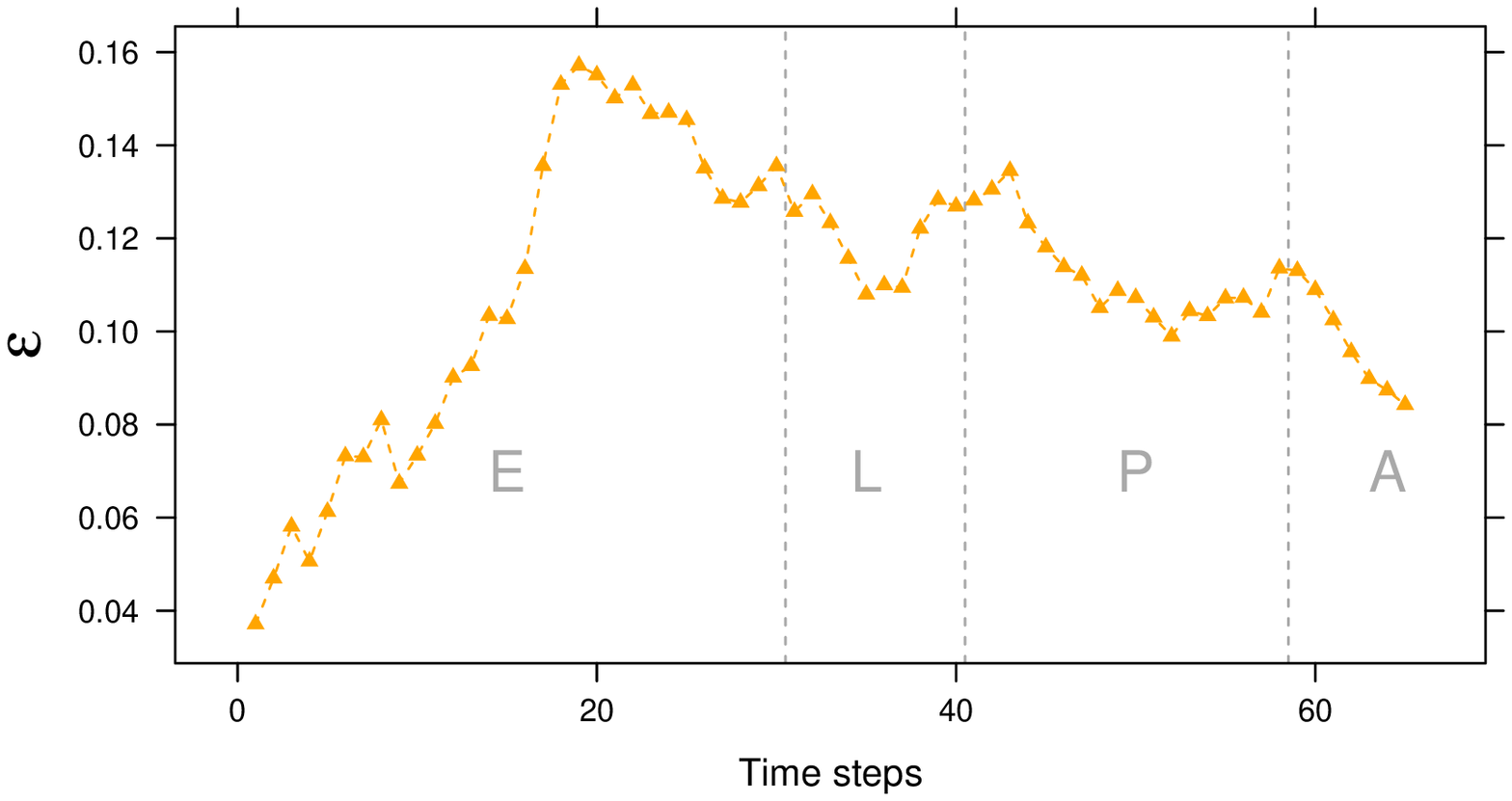}  \\
(a) $\epsilon(N_i,N_{i+1})$ & (b) $\epsilon(N_i,N_1)$ \\
\includegraphics[width=0.49\textwidth]{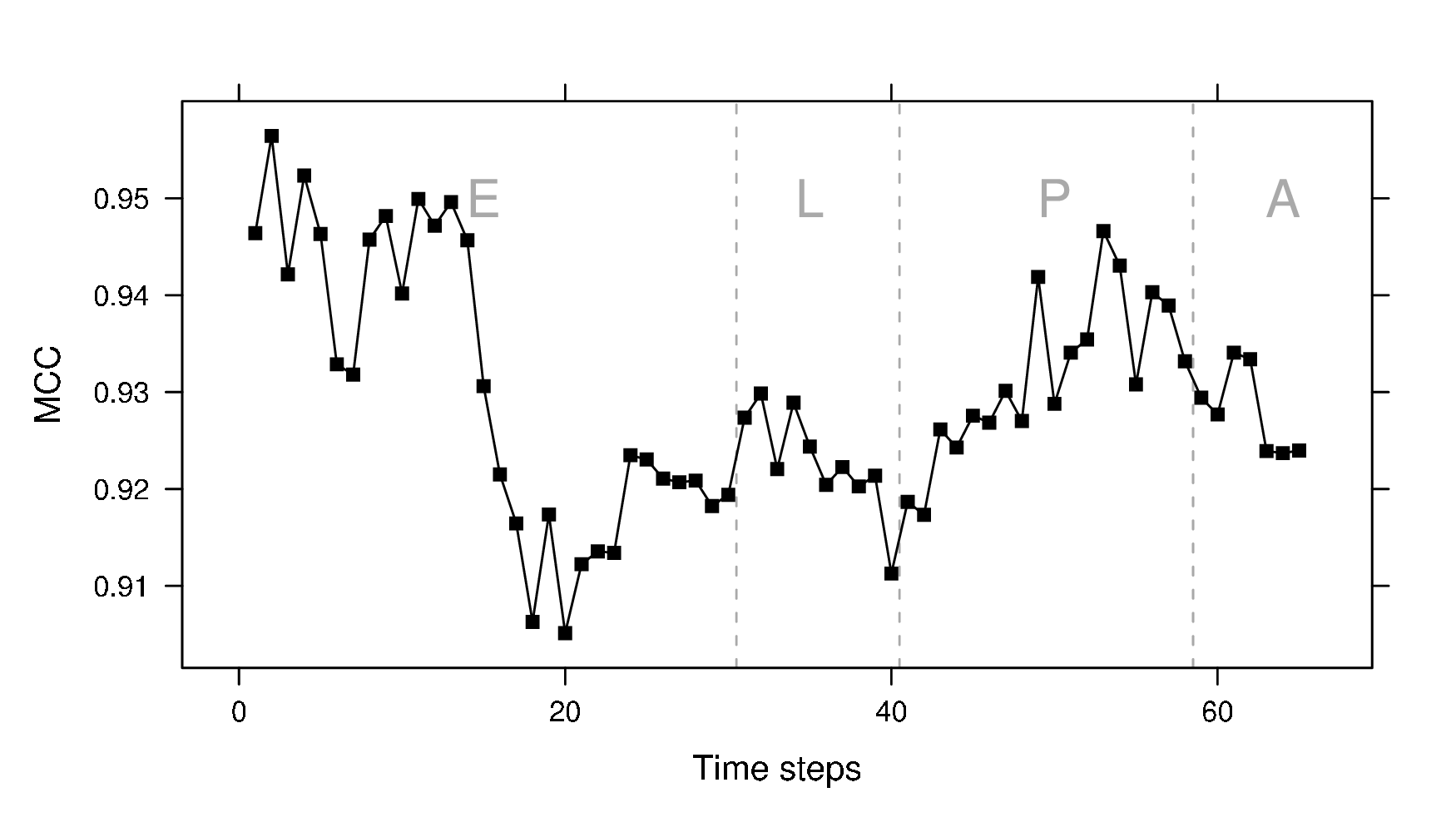} & \includegraphics[width=0.49\textwidth]{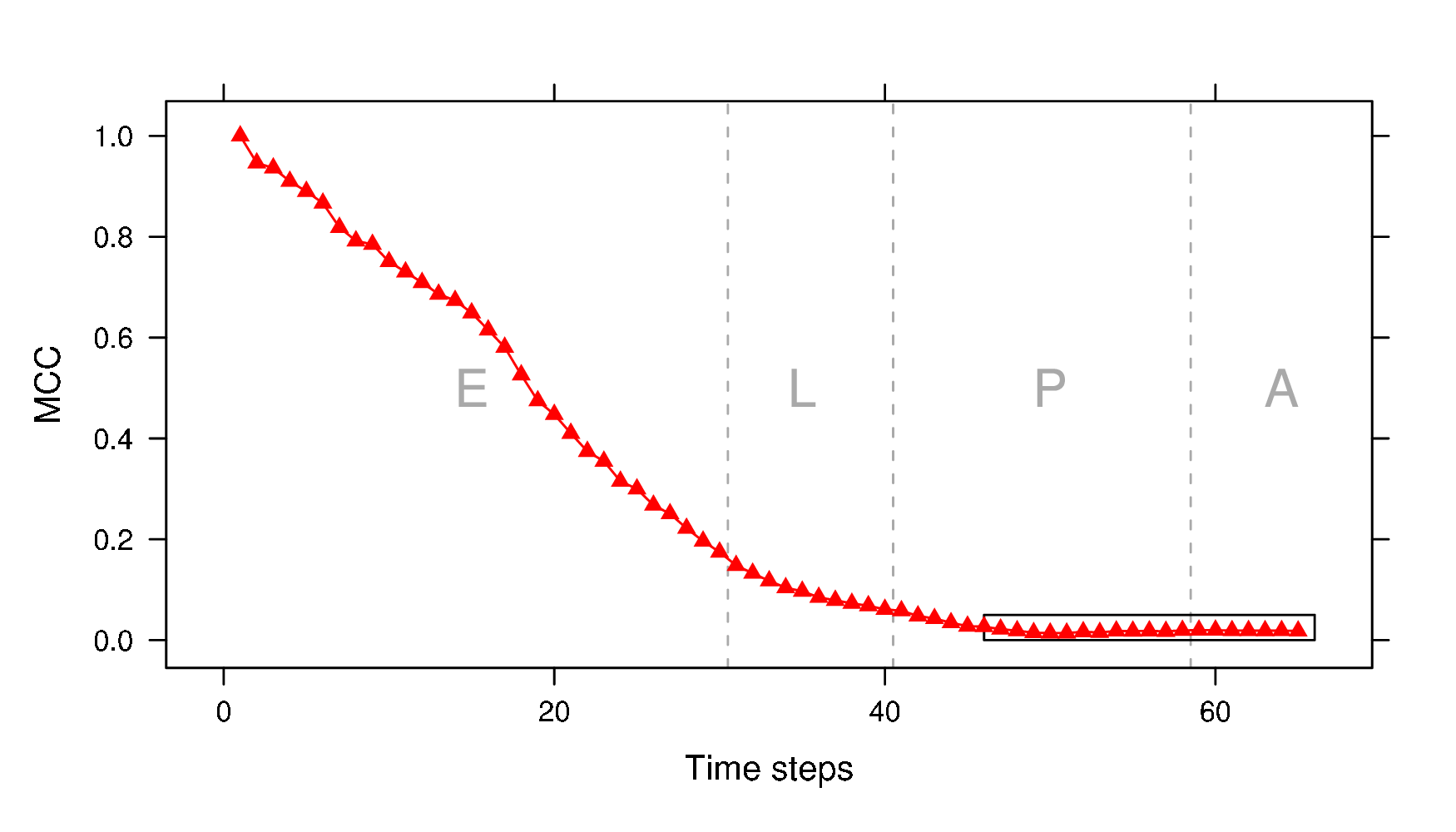}  \\
(c) $MCC(N_i,N_{i+1})$ & (d) $MCC(N_i,N_1)$ \\
 & \includegraphics[width=0.49\textwidth]{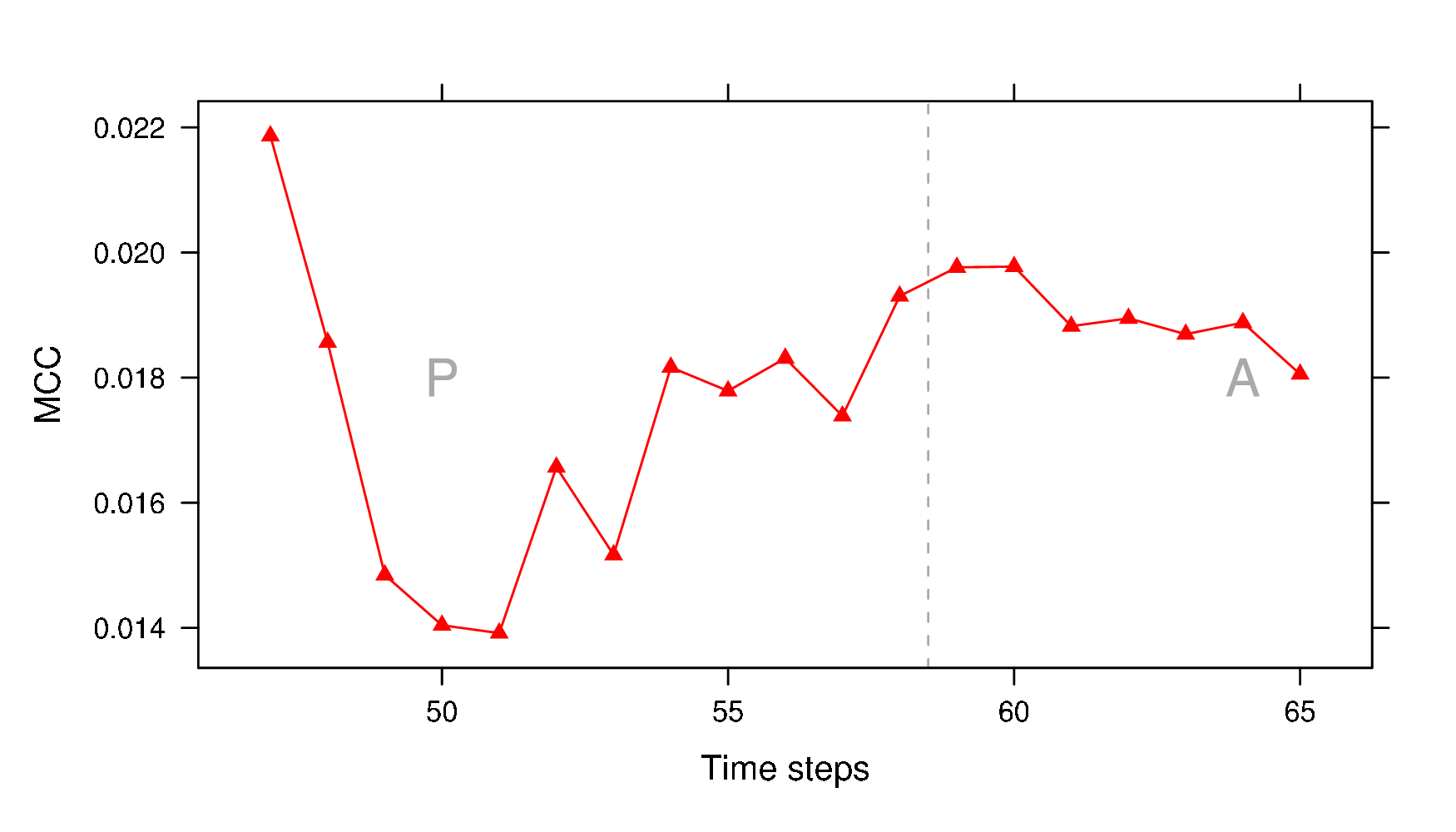}\\
& (e) $MCC(N_{i\geq 47},N_1)$ 
\end{tabular}
\end{center}
\caption{$\epsilon$ and MCC distances computed on consecutive elements (a and c) and with respect to the first element of the series (b and d) of the dynamic transcriptional network on 66 time points of the developmental cycle of \textit{D. melanogaster}; in (e), zoom of (d) on the time points 47-65.}
\label{fig:time}
\end{figure}

In \citep{kolar10estimating}, the authors used the Keller algorithm to infer the gene regulatory networks of \textit{Drosophila melanogaster} from a time series of gene expression data measured during its full life cycle.
They selected 66 time points during the developmental cycle, spanning across four different stages (Embryonic -- time points 1-30, Larval -- t.p. 31-40, Pupal -- t.p. 41-58, Adult -- t.p. 59-66), following the dynamics of 588 gene ontological groups and then constructing a time series of inferred networks $N_i$\footnote{Adjacency matrices are available at \url{http://cogito-b.ml.cmu.edu/keller/downloads.html}}.
Hereafter we evaluate the structural differences between $N_i$ and $N_{i+1}$ and the distance between $N_i$ and the initial network $N_1$, measured either by the Ipsen-Mikhailov distance or by MCC: the resulting plots are displayed in Fig.~\ref{fig:time}.
The largest variations, both between consecutive terms and with respect to the initial network $N_1$, occur in the embrional stage (E). 
As expected, the variations between consecutive terms (panels (a) and (c)) are smaller, while more relevant changes occur comparing a term with $N_1$.
In particular, it is interesting to note that the dynamics of the networks move $N_i$ away from $N_1$ until time points 20, then the following terms start getting closer again to $N_1$ in terms of Ipsen-Mikhailov distance.
The same trend is captured by MCC, but with lower resolution: the fact that MCC curve is ascending from its minimum in the last 15 time points can be appreciated only by zooming in from panel (d) to panel (e).
This means that, after the embrional stage, the network is getting structurally more and more similar again to $N_1$, but with a limited number of links matching those of $N_1$.

\subsection*{Networks in profiling tasks}
In the papers \citep{budhu08identification,ji09microrna}, the authors introduced and analyzed a dataset\footnote{Available at Gene Expression Omnibus (GEO) \url{http://www.ncbi.nlm.nih.gov/geo/} at the accession number GSE6857} collecting 482 tissue samples from 241 patients affected by hepatocellular carcinoma (HCC). 
For each patients, a sample from cancerous hepatic tissue and a sample from surrounding non-cancerous hepatic tissue have been hybridized on the Ohio State University CCC MicroRNA Microarray Version 2.0 platform consisting of 11520 probes collecting expressions of 250 non-redundant human and 200 mouse microRNA (miRNA).

\begin{table}[!ht]
\caption{Workflow of the machine learning pipeline for the profiling tasks on the $\mathcal{HCC}$ dataset.}
\label{tab:workflow}
\begin{center}
\tiny{
\texttt{
\begin{enumerate}
\item Preprocessing phase: imputation of missing values \citep{troyanskaya01missing} and discarding probes corresponding to non-human (mouse and controls) miRNA;
\item Obtaining a dataset $\mathcal{HCC}$ of 240+240 paired samples described by 210 human miRNA
\item Three profiling experiments: discriminate the two classes Tumoral (T) and non Tumoral (nT) within the whole set $\mathcal{HCC}$, in the subset $\mathcal{HCC}_M$ of the 210+210 samples belonging to male patients (M) and in the subset $\mathcal{HCC}_F$ of the 30+30 samples belonging to female (F) patients;
\item Data Analysis Protocol (DAP) as in \citep{budhu08identification}: 1000 $\times$ 10-fold Cross Validation;
\item Classifier: Spectral Regression Discriminant Analysis (SRDA) \citep{cai08srda}, $\alpha=100$, Feature Ranking: Entropy-based Recursive Feature Elimination (E-RFE) \citep{furlanello03entropy};
\item Performance: MCC averaged on the 1000 test set for models with different number of features; confidence intervals are computed as 95\% student bootstrap;
\item Ranked list Stability: the Canberra stability indicator $I$ \citep{jurman08algebraic} defined as the mean of mutual Canberra distances among the lists, normalized with respect to the whole set of possible permutations.
The smaller the indicator value, the higher the stability level of the lists, with 0 corresponding to a set of 10000 identical lists and 1 to a set of randomly ranked lists;
\item Results: The model with 20 features is a reasonable compromise between classifier performance, list stability and small number of features: for the tasks with all samples, $MCC=0.845$ $CI=(0.839,0.850)$, $I=0.166$, while for the other two cases the analogous values are $M=(0.931,(0.927,0.934),0.323)$ and $F=(0.859,(0.846,0.871),0.349)$;
\item Optimal list: for each of the three problems is computed as the top-20 sublist of the whole Borda list \citep{jurman08algebraic, borda81memoire};
\item In Tab.~\ref{tab:common} we list the 16 miRNA common to at least two out of the three top-20 models: in particular, 7 miRNA are common to all the three problems.
\end{enumerate}
}
}
\end{center}
\end{table}

By the Machine Learning pipeline detailed in Tab.~\ref{tab:workflow} we extract the top-20 optimal set of features discriminating cancer samples from controls. 
Most of them are already known in literature as associated with hepatocellular carcinoma.

\begin{table}[!ht]
\caption{Common miRNA}
\label{tab:common}
\begin{center}
\normalsize{
\begin{tabular}{c|cc}
\hline
$\mathcal{HCC}$ &  hsa-mir-021-prec-17No1  & hsa-mir-099-prec-21 \\
& hsa-mir-128b-precNo1& hsa-mir-21No1\\
$\mathcal{HCC}_{M,F}$ & hsa-mir-221-prec & hsa-mir-222-precNo1\\
&  hsa-mir-26a-1No2\\
\hline
$\mathcal{HCC}$ &   hsa-mir-122a-prec \\
$\mathcal{HCC}_{M}$&\\
\hline
$\mathcal{HCC}$ & hsa-mir-100No1&  hsa-mir-125b-1\\
& hsa-mir-199b-precNo2\\
$\mathcal{HCC}_{F}$ & hsa-mir-219-1No2 & hsa-mir-222-precNo2 \\
\hline
$\mathcal{HCC}_{M}$ & hsa-mir-130a-precNo2 \\
$\mathcal{HCC}_{F}$ & hsa-mir-146-prec \\
\hline
\end{tabular}
}
\end{center}
\end{table}

\begin{figure}[!ht]
\begin{center}
\begin{tabular}{ccc}
\includegraphics[width=0.4\textwidth]{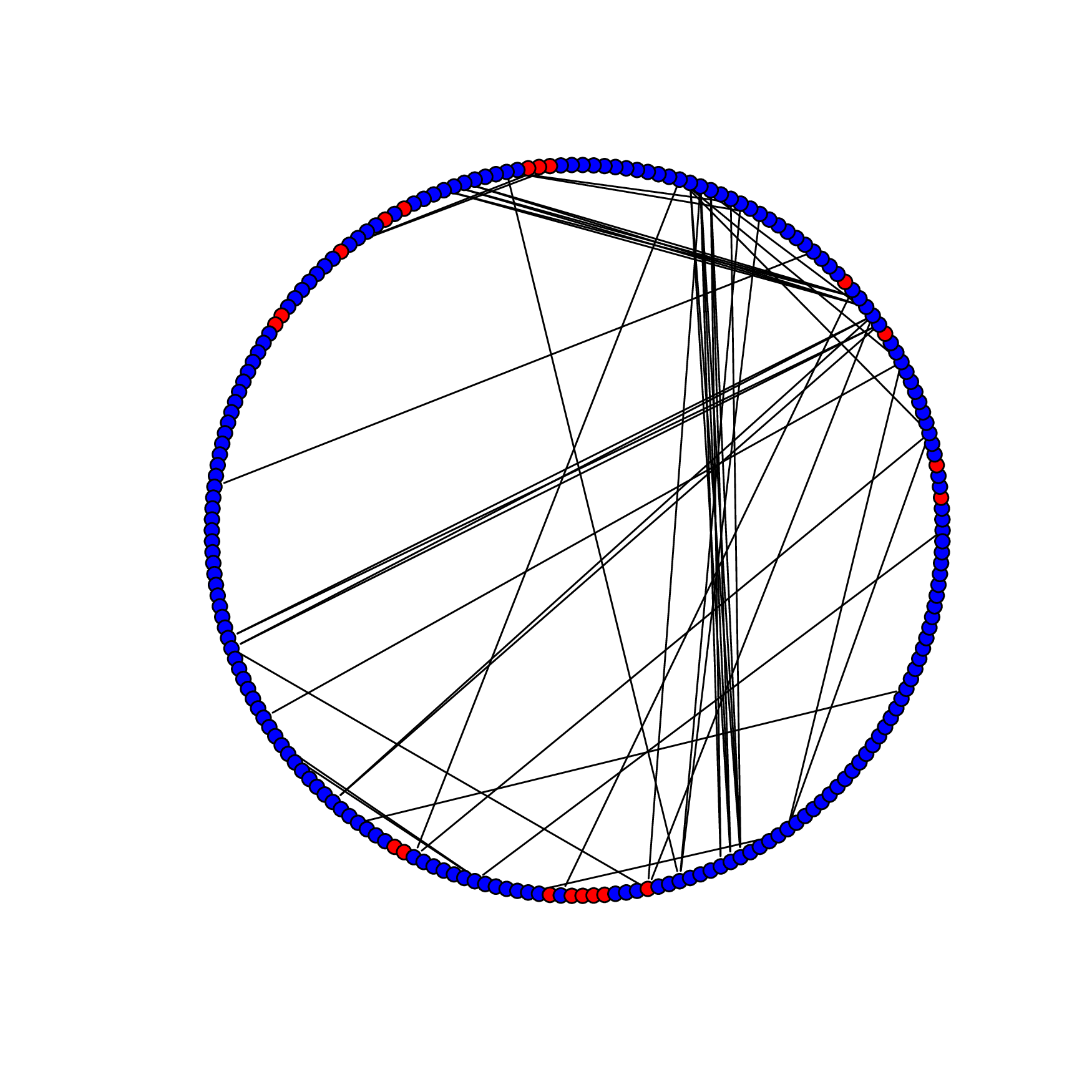} &\phantom{AAAAAA} & \includegraphics[width=0.4\textwidth]{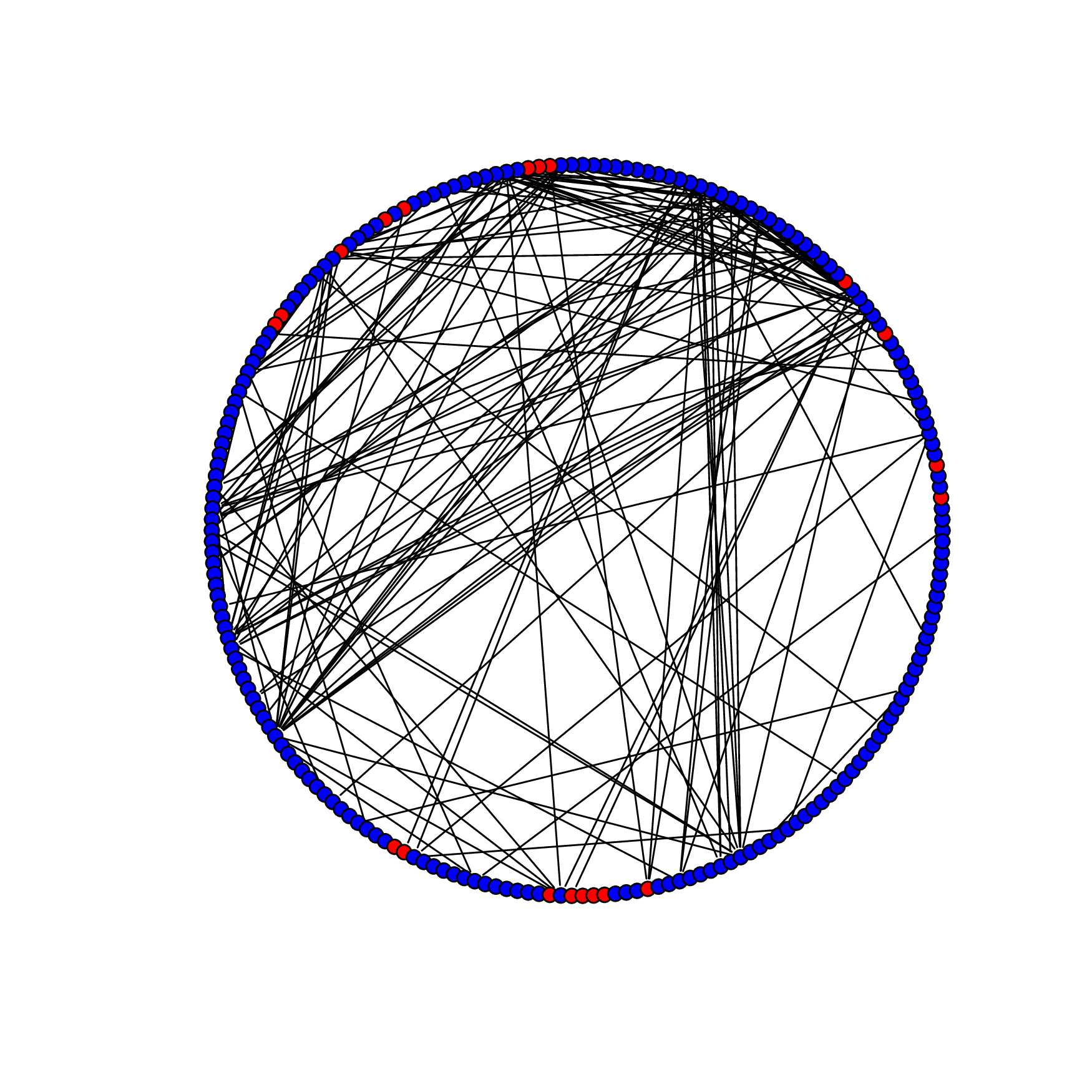} \\
(a) (M+F) T & & (b) (M+F) nT \\
\includegraphics[width=0.4\textwidth]{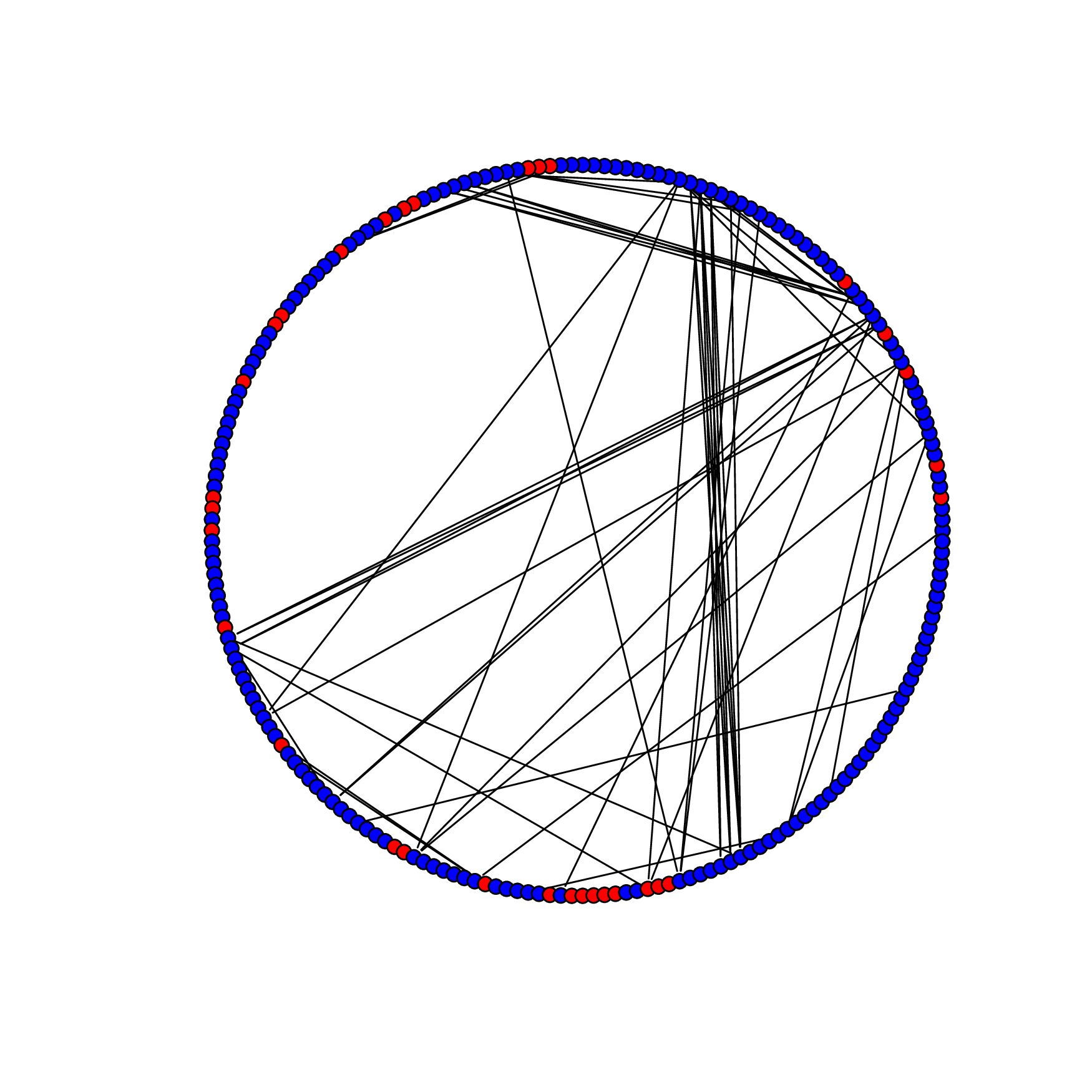} & & \includegraphics[width=0.4\textwidth]{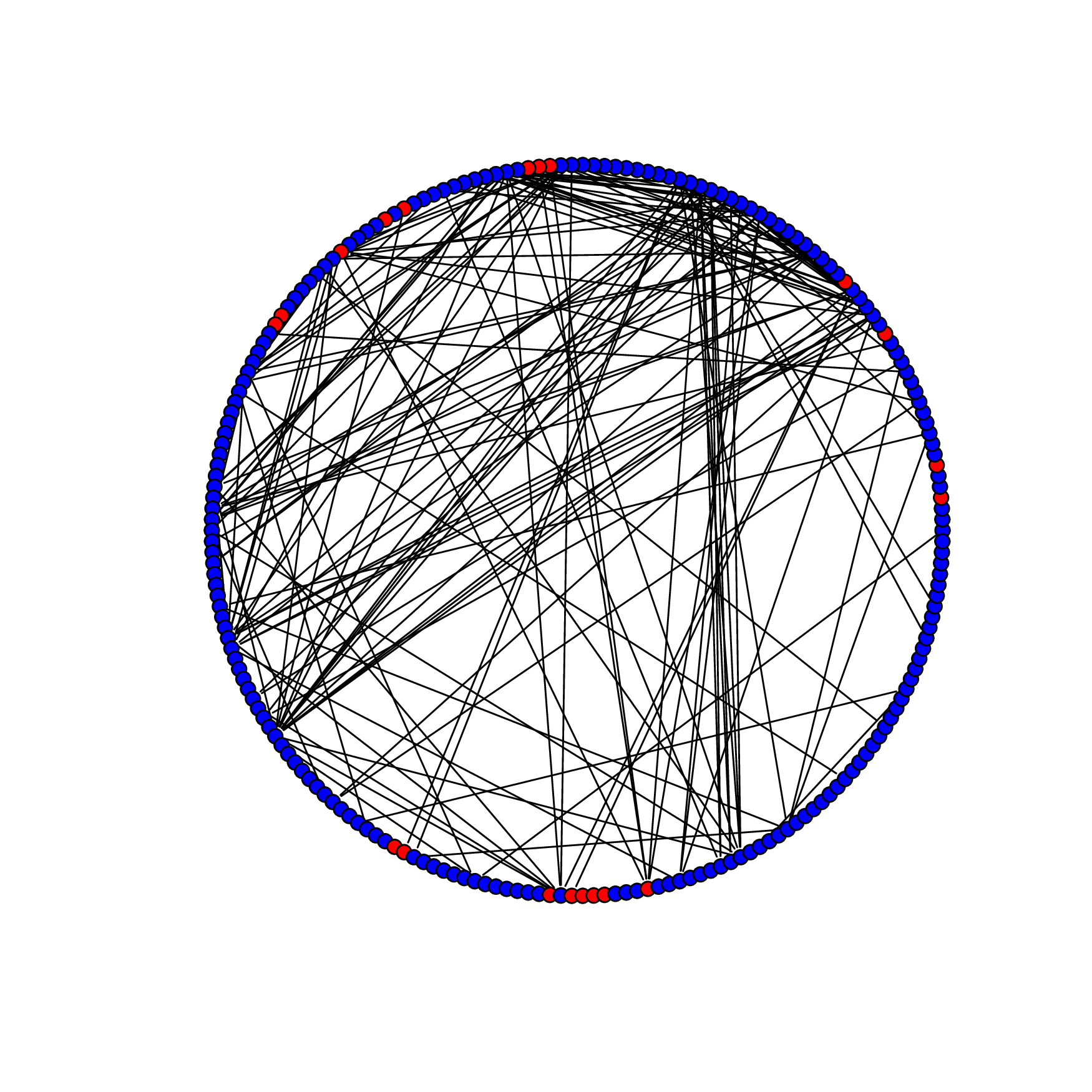} \\
(c) M T &&  (d) M nT \\
\includegraphics[width=0.4\textwidth]{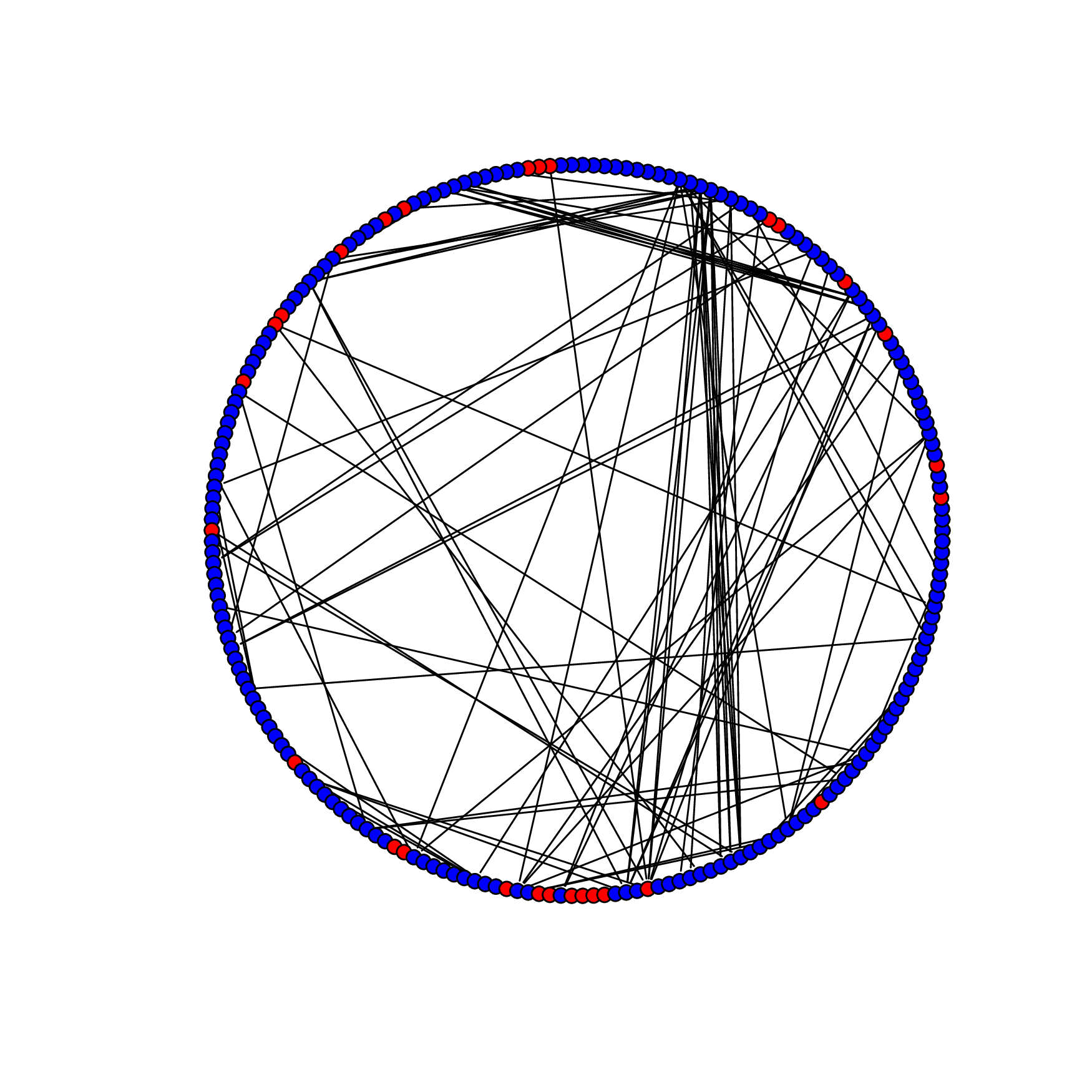} & & \includegraphics[width=0.4\textwidth]{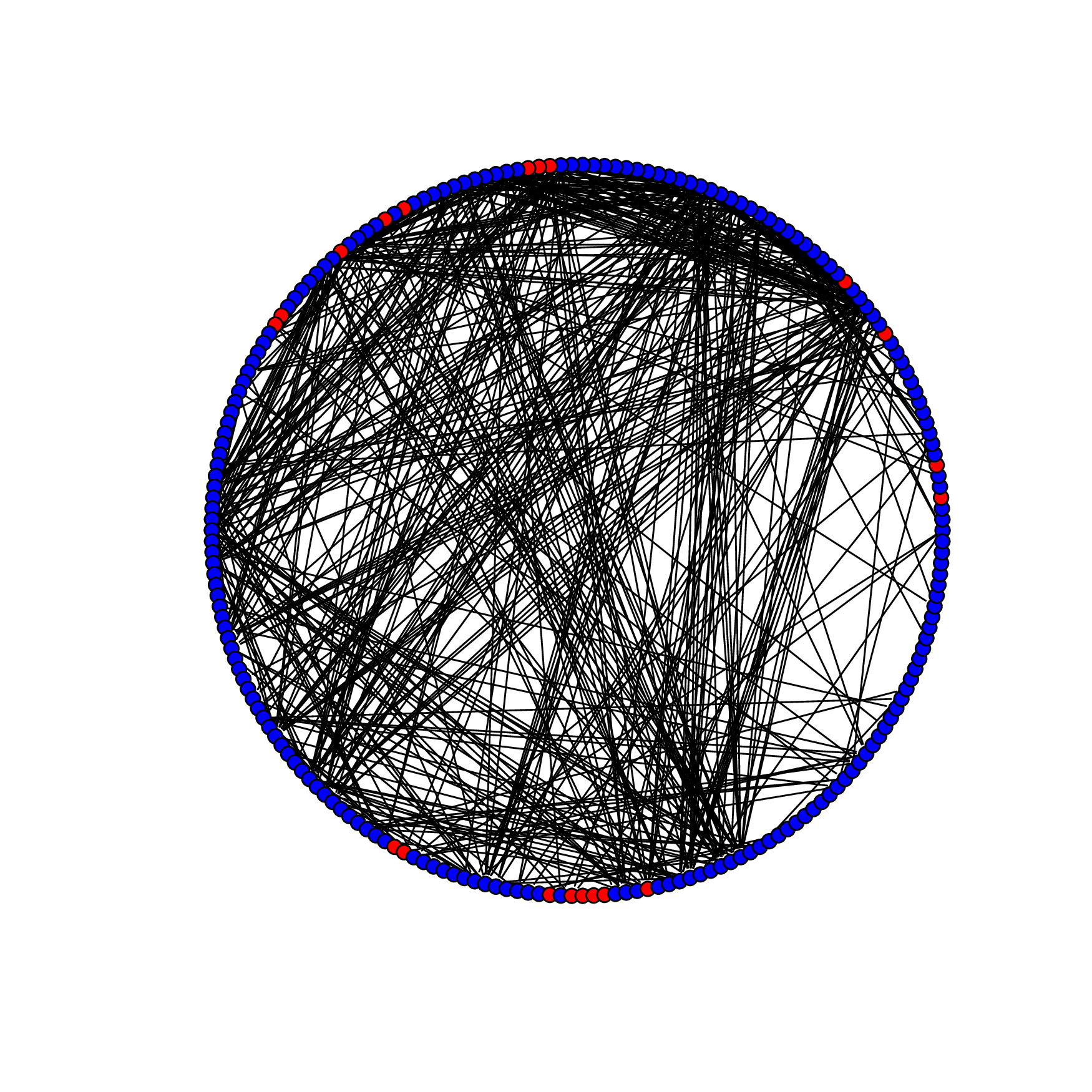} \\
(e) F T & & (f) F nT \\
\end{tabular}
\end{center}
\caption{Relevance network at correlation 0.85 for (left) the tumoral samples (T) and (right) the control tissues (nT), for the whole dataset (M+F), the male patients (M) and the female patients (F), with the top-20 ranked features marked as red nodes.}
\label{fig:mirna}
\end{figure}

The following phase consists in the construction of the six weighted miRNA networks associated to the data subsets MT, MnT, FT, FnT, (M+F)T, (M+F)nT by using three different inference algorithm: WGCNA \citep{zhang05general,zhao10weighted,horvath11weighted}, Aracne \citep{margolin06aracne} and CLR \citep{faith07large}, also considering their binarized versions after thresholding.
As an example, in Fig.~\ref{fig:mirna} we show the correlation networks at threshold $0.85$ in all the six considered cases: the number of links in the healthy tissue case is always larger than in the cancerous tissue case.
Using Ipsen-Mikhailov distance, it is now possibile to quantitatively explore the similarity among the miRNA profile networks: in what follows, we show some examples.

\begin{figure}[!ht]
\includegraphics[width=\textwidth]{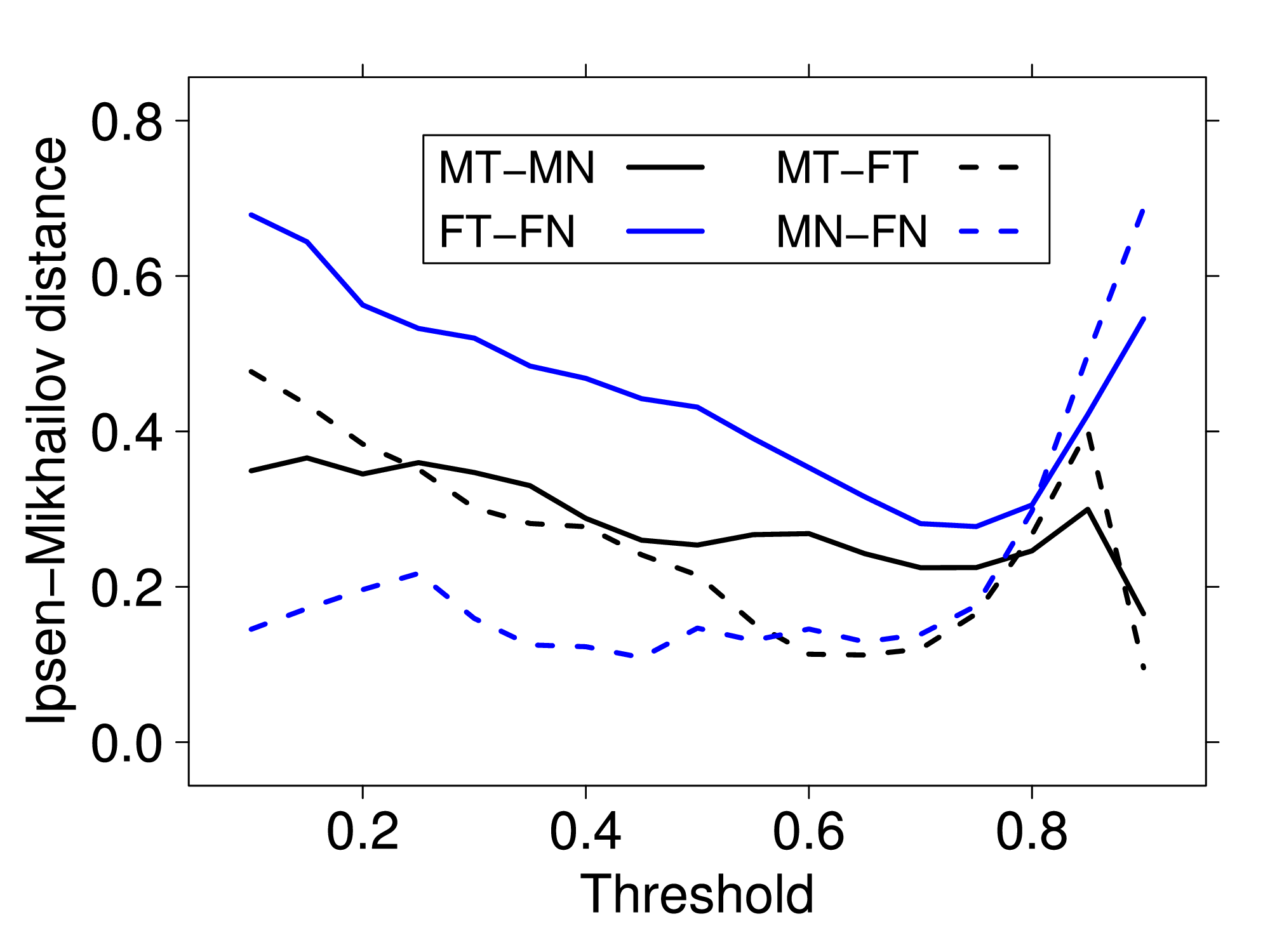}
\caption{Evolutions of distances between 4 couples of correlation networks as a function of the correlation threshold.}
\label{fig:evolving}
\end{figure}

For instance, in Fig.~\ref{fig:evolving} we show how distances between four couples of correlation networks evolve with the correlation threshold travelling between 0.1 and 0.9.
The two closest networks are those corresponding to the Control tissues, with a classwise related trend independent from gender. 
In fact, the two curves expressing respectively the distance in the Tumorous tissue case between Male and Female patients and the corresponding curve for the Control tissue have a similar shape up to correlation threshold 0.8.
Finally, for Female patients, the Tumoral network is quite distant from the Control one, hilighting a wider biological transformation caused by the disease than in Male patients.

\begin{figure}[!ht]
\begin{center}
\begin{tabular}{cc}
\includegraphics[width=0.4\textwidth]{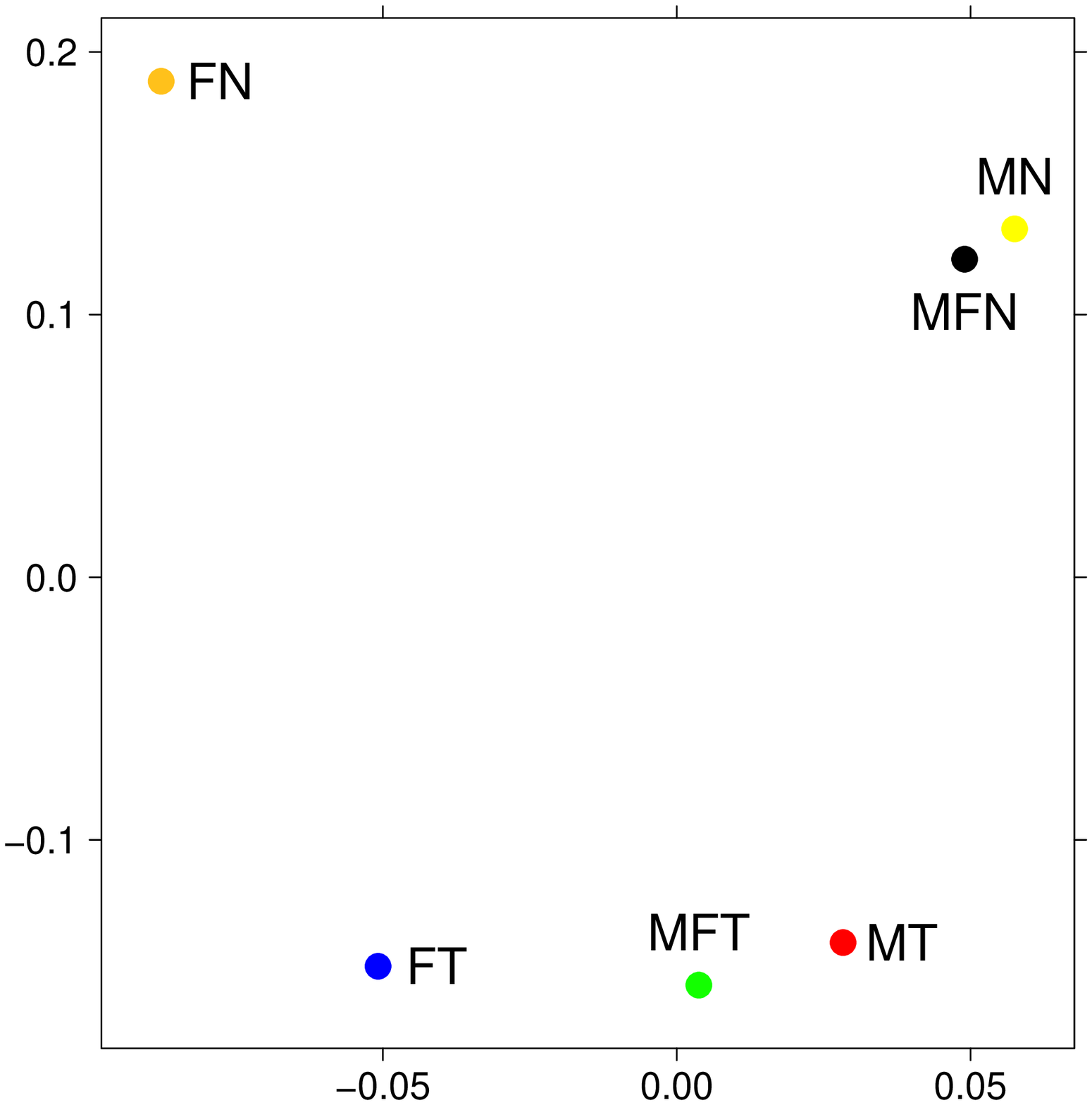} & \includegraphics[width=0.4\textwidth]{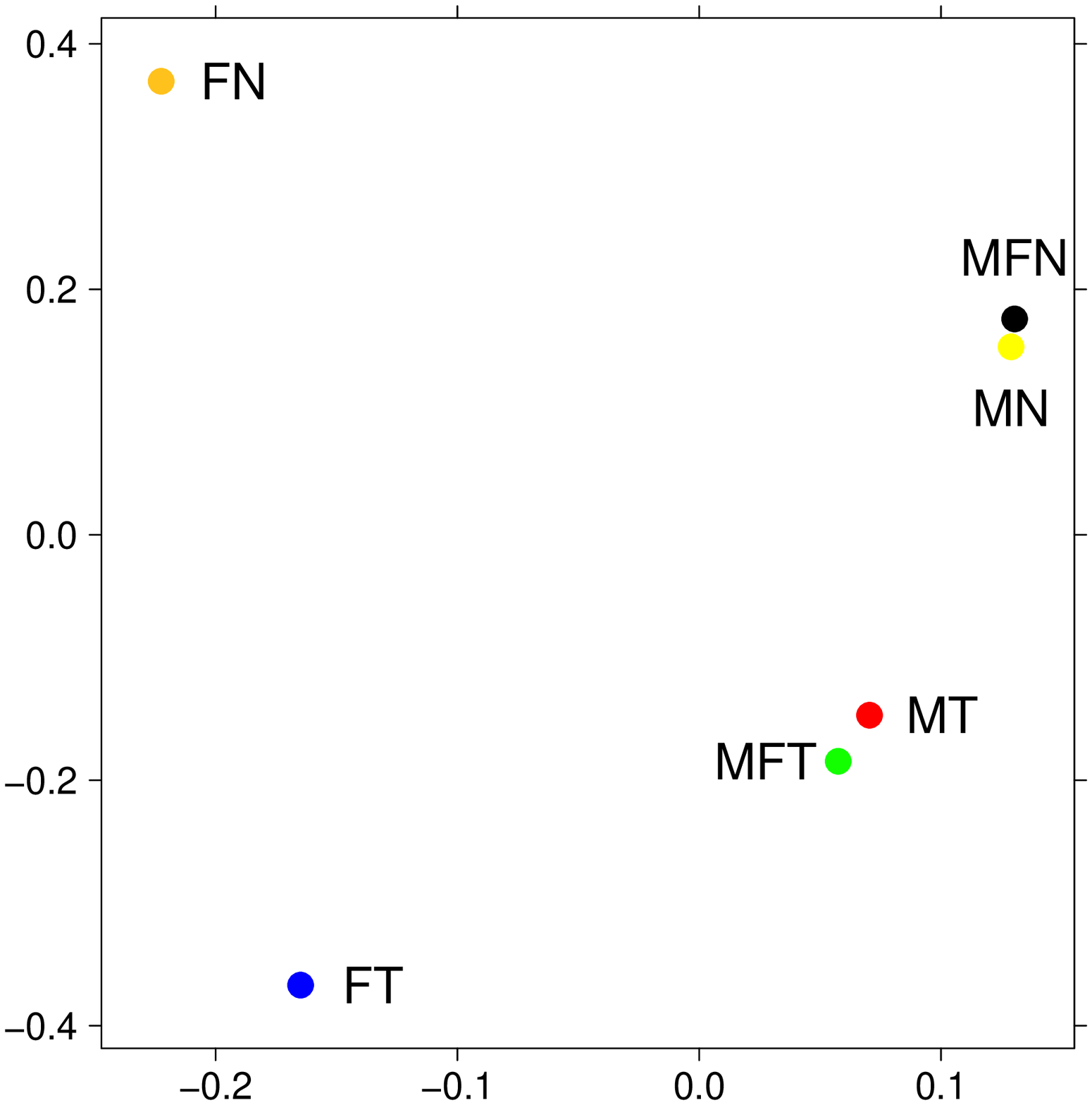} \\
\multicolumn{2}{c}{Weighted WGCNA networks} \\
\includegraphics[width=0.4\textwidth]{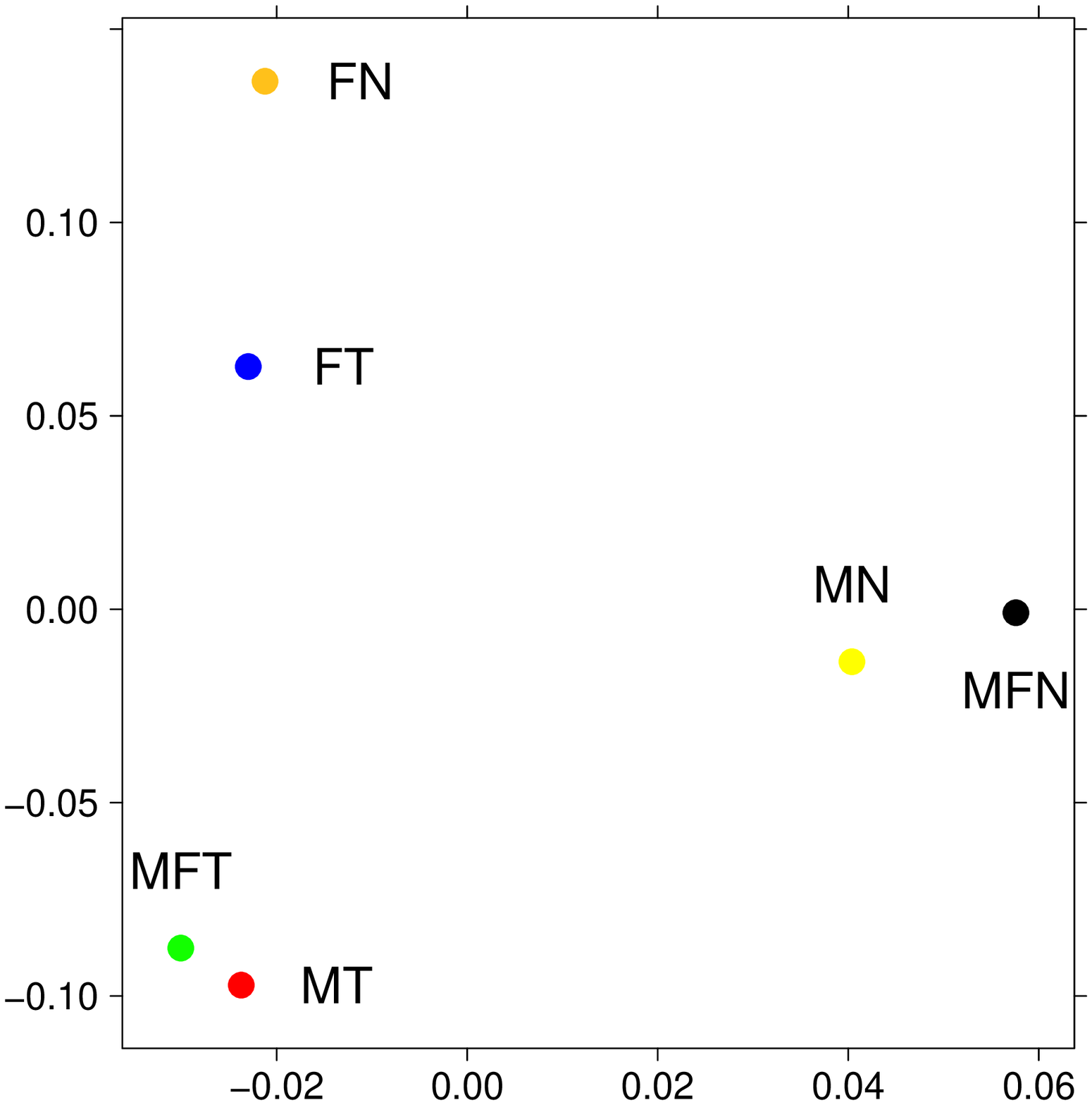} & \includegraphics[width=0.4\textwidth]{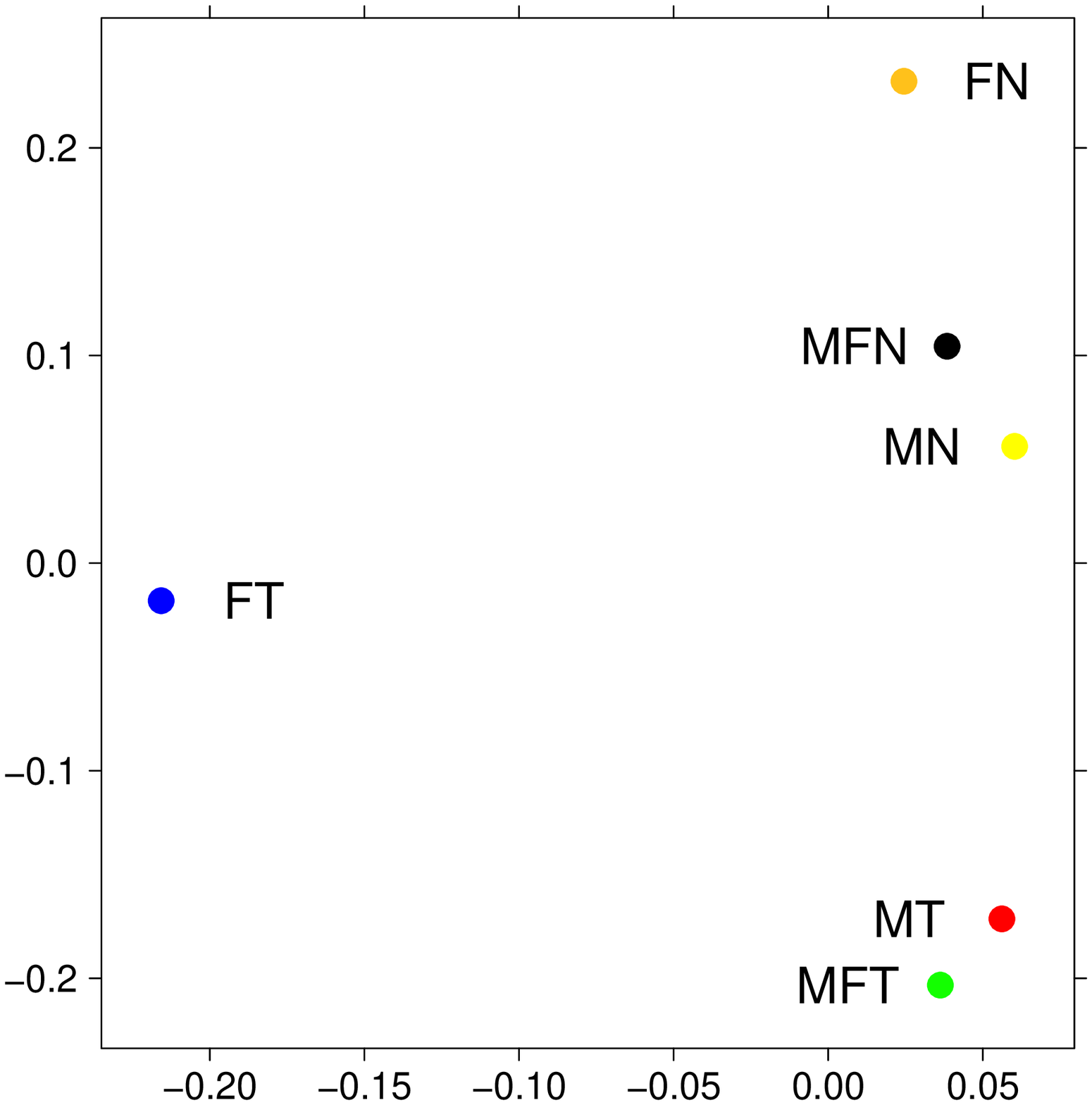} \\
\multicolumn{2}{c}{Weighted Aracne networks} \\
\includegraphics[width=0.4\textwidth]{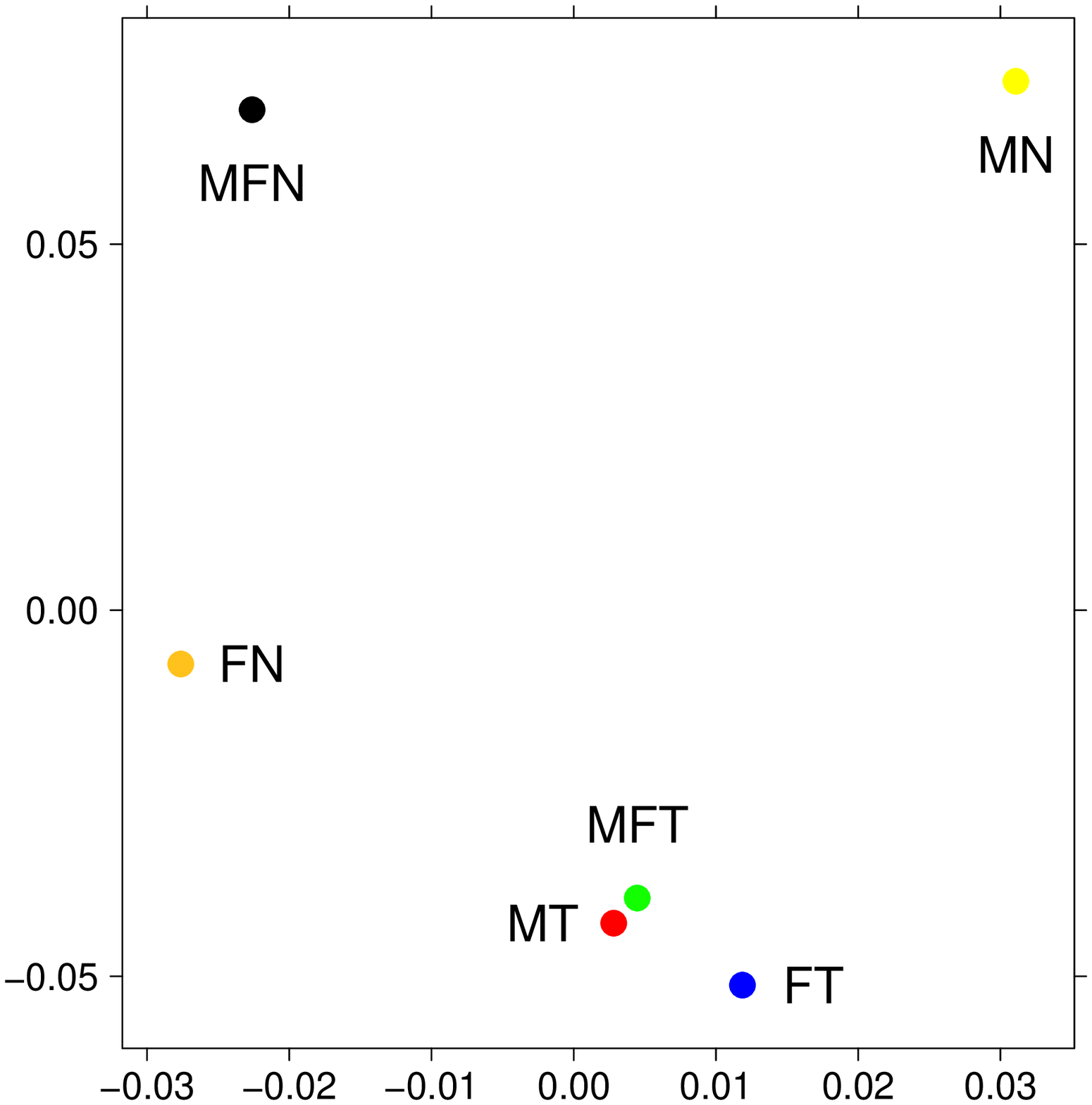} & \includegraphics[width=0.4\textwidth]{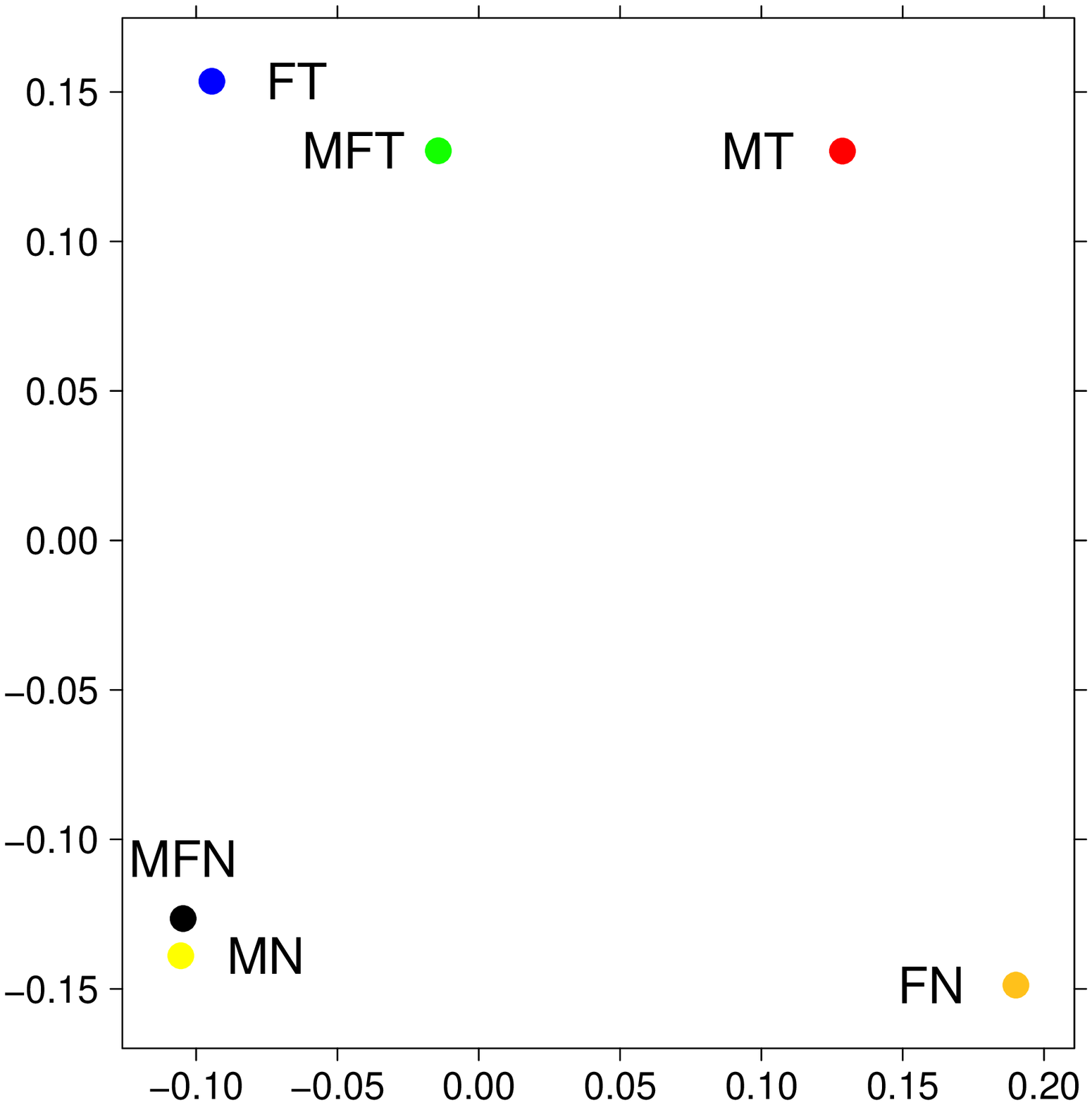} \\
\multicolumn{2}{c}{Weighted CLR networks} \\
\end{tabular}
\end{center}
\caption{Multidimensional scaling of the distances listed in Tab.~\ref{tab:d2}, for the complete set of miRNA (left panels) and the top-20 subset (right panels).}
\label{fig:mirnamds}
\end{figure}

In Tab. \ref{tab:d2} the Ipsen-Mikhailov distances are reported among all six weighted networks for different methods, either on the whole set of 210 miRNA or on the top-20 set of optimal features. 
The corresponding multidimensional scaling projections are displayed in Fig.~\ref{fig:mirnamds}.
\begin{table}[!ht]
\caption{Ipsen-Mikhailov distances among all six networks for different inference methods, on the whole set of 210 miRNA (upper triangular) or on the top-20 set of optimal features (lower triangular).}
\label{tab:d2}
\begin{center}
\normalsize{
\begin{tabular}{c|cccccc}
\hline
\multicolumn{7}{c}{Mutual distances for weighted WGCNA networks}\\ 
\hline
 	& F T 	& M T 	& M+F T 	& F nT 	& M nT 	& M+F nT \\ 
F T &  & 0.1440 & 0.1228 & 0.3538 & 0.3056 & 0.2929 \\ 
M T & 0.4091 &  & 0.0838 & 0.3498 & 0.2845 & 0.2742 \\ 
M+F T & 0.3996 & 0.1091 &  & 0.3587 & 0.3012 & 0.2871 \\ 
F nT & 0.7648 & 0.5980 & 0.6272 &  & 0.1659 & 0.1634 \\ 
M nT & 0.5998 & 0.3687 & 0.4176 & 0.4403 &  & 0.0500 \\ 
M+F nT & 0.6197 & 0.3962 & 0.4389 & 0.4344 & 0.0594 &  \\ 
\hline
\multicolumn{7}{c}{Mutual distances for weighted Aracne networks}\\ 
\hline
 	& F T 	& M T 	& M+F T 	& F nT 	& M nT 	& M+F nT \\ 
\hline
F T &  & 0.1764 & 0.1636 & 0.1219 & 0.1210 & 0.1179 \\ 
M T & 0.3162 &  & 0.0408 & 0.2358 & 0.1132 & 0.1299 \\ 
M+F T & 0.3237 & 0.1241 &  & 0.2280 & 0.1075 & 0.1271 \\ 
F nT& 0.3604 & 0.4183 & 0.4400 &  & 0.1685 & 0.1658 \\ 
M nT& 0.2934 & 0.2454 & 0.2926 & 0.2344 &  & 0.0570 \\ 
M+F nT& 0.2945 & 0.2960 & 0.3370 & 0.2120 & 0.1194 &  \\ 
\multicolumn{7}{c}{Mutual distances for weighted CLR networks} \\
\hline
 	& F T 	& M T 	& M+F T 	& F nT 	& M nT 	& M+F nT \\ 
\hline
F T &  & 0.0043 & 0.0037 & 0.0056 & 0.1194 & 0.1233 \\ 
M T & 0.3260 &  & 0.0008 & 0.0030 & 0.1158 & 0.1102 \\ 
M+F T & 0.2441 & 0.2506 &  & 0.0028 & 0.1123 & 0.1067 \\ 
F nT& 0.4223 & 0.3571 & 0.3788 &  & 0.0964 & 0.0017 \\ 
M nT& 0.3380 & 0.3669 & 0.3160 & 0.3235 &  & 0.0011 \\ 
M+F nT& 0.3318 & 0.3577 & 0.3012 & 0.3251 &  &  \\ 
\end{tabular}
}
\end{center}
\end{table}

The distances among networks show that there are substantial differences not only between the Tumorous/NonTumorous tissue samples, but also between Male and Female patients, both on the cancerous and the surrounding healthy tissue relevance networks.
In particular, it can be pointed out that the networks corresponding to the tumoral tissue for female patients has a deeply different structure with respect to all other networks, while the differences between the models on all patients and those on the sole male population are smaller. 
This may be an effect of the different numerosity between male and female patients (210 versus 30) for WGCNA networks, but it is confirmed also by the Aracne algorithm which is less sensible to sample size differences.
Finally, CLR is the algorithm where the difference between the networks built on the whole set of features and those built on the top-20 subset are more relevant.

We can conclude with an analysis of distances across inference methods for networks associated to a given sample subset, listed in Tab.~\ref{tab:across_methods}.
The structures inferred by WGCNA and CLR are the closest when the full set of features are used, but distances are small among methods.
The situation radically changes when the subset of optimal features is considered: in this case, WGCNA and Aracne tend to build up very similar networks for all the sample subsets, while CLR is going astray, confirming the observations drawn from the multidimensional scaling plots of Fig.~\ref{fig:mirnamds}.
\begin{table}[ht]
\caption{Ipsen-Mikhailov distances between weighted network inferred from the same sample subsets, for different inference methods WGCNA (W), Aracne (A) and CLR (C), for the whole set of miRNA (upper half) and the optimal top-20 miRNA subset (lower half).}
\label{tab:across_methods}
\begin{center}
\begin{tabular}{c|ccc}
\hline
\multicolumn{4}{c}{Full set of 210 miRNA}\\
\hline
& W,A & W,C & A,C \\ 
\hline
F T & 0.6693 & 0.5732 & 0.7164 \\ 
M T & 0.6860 & 0.5648 & 0.7176 \\ 
MF T & 0.6831 & 0.5637 & 0.7207 \\ 
F nT & 0.6821 & 0.5289 & 0.7012 \\ 
M nT & 0.6678 & 0.5360 & 0.7011 \\ 
M+F nT & 0.6650 & 0.5337 & 0.6998 \\ 
\hline
\multicolumn{4}{c}{Optimal subset of 20 miRNA}\\
\hline
& W,A & W,C & A,C \\ 
\hline
F T & 0.2537 & 0.9567 & 0.9365 \\ 
M T & 0.3557 & 0.8320 & 0.8984 \\ 
M+F T & 0.3807 & 0.8343 & 0.9103 \\ 
F nT& 0.5192 & 0.8335 & 0.8625 \\ 
M nT& 0.3707 & 0.8198 & 0.8562 \\ 
M+F nT& 0.3628 & 0.8192 & 0.8555 \\ 
\hline
\end{tabular}
\end{center}
\end{table}
 
\subsection*{Subset stability in network reconstruction}
\begin{figure}[!ht]
\begin{center}
\begin{tabular}{cc}
\includegraphics[width=0.45\textwidth]{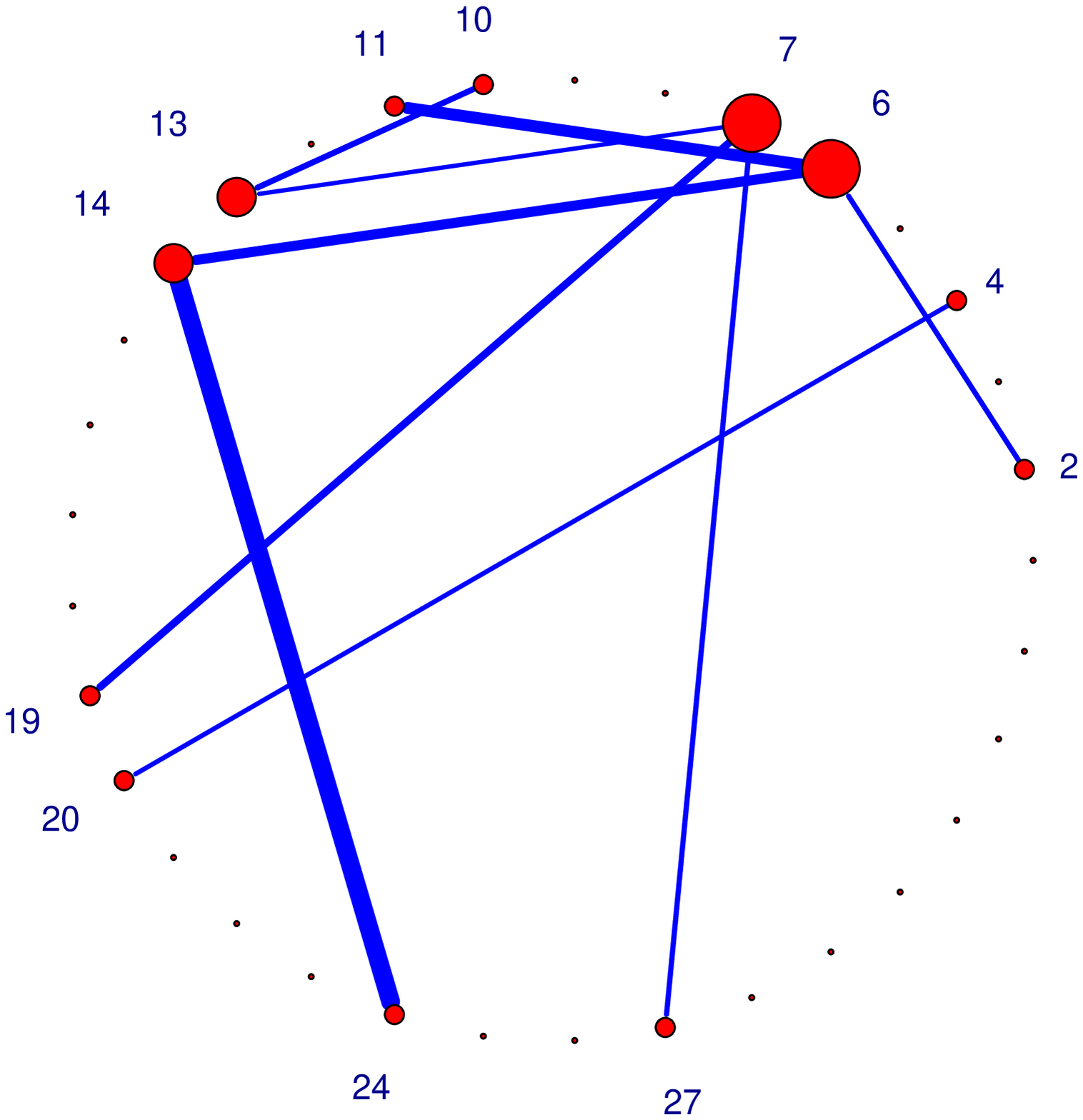} &
\includegraphics[width=0.45\textwidth]{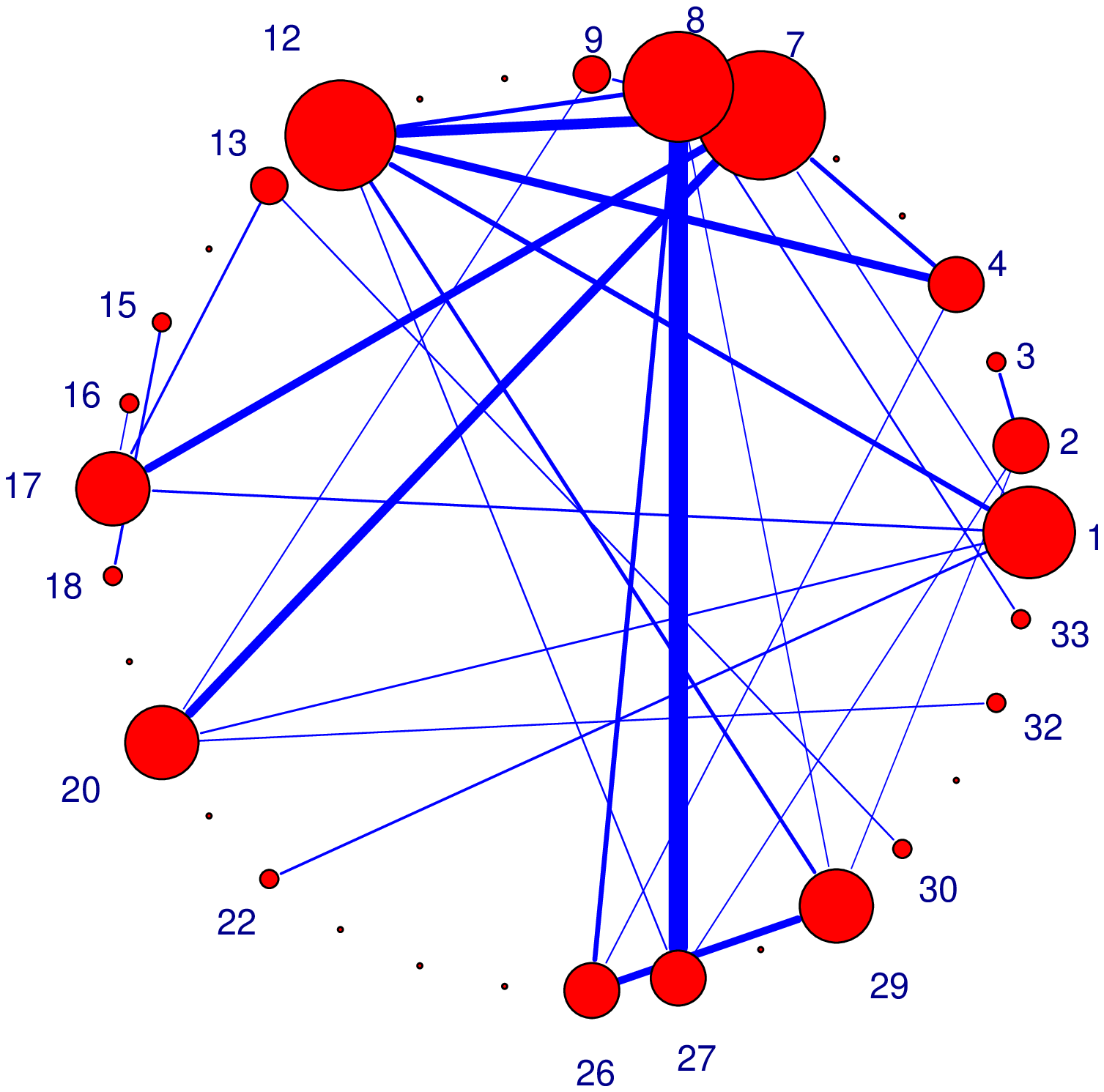}\\
(a) & (b) \\
\includegraphics[width=0.45\textwidth]{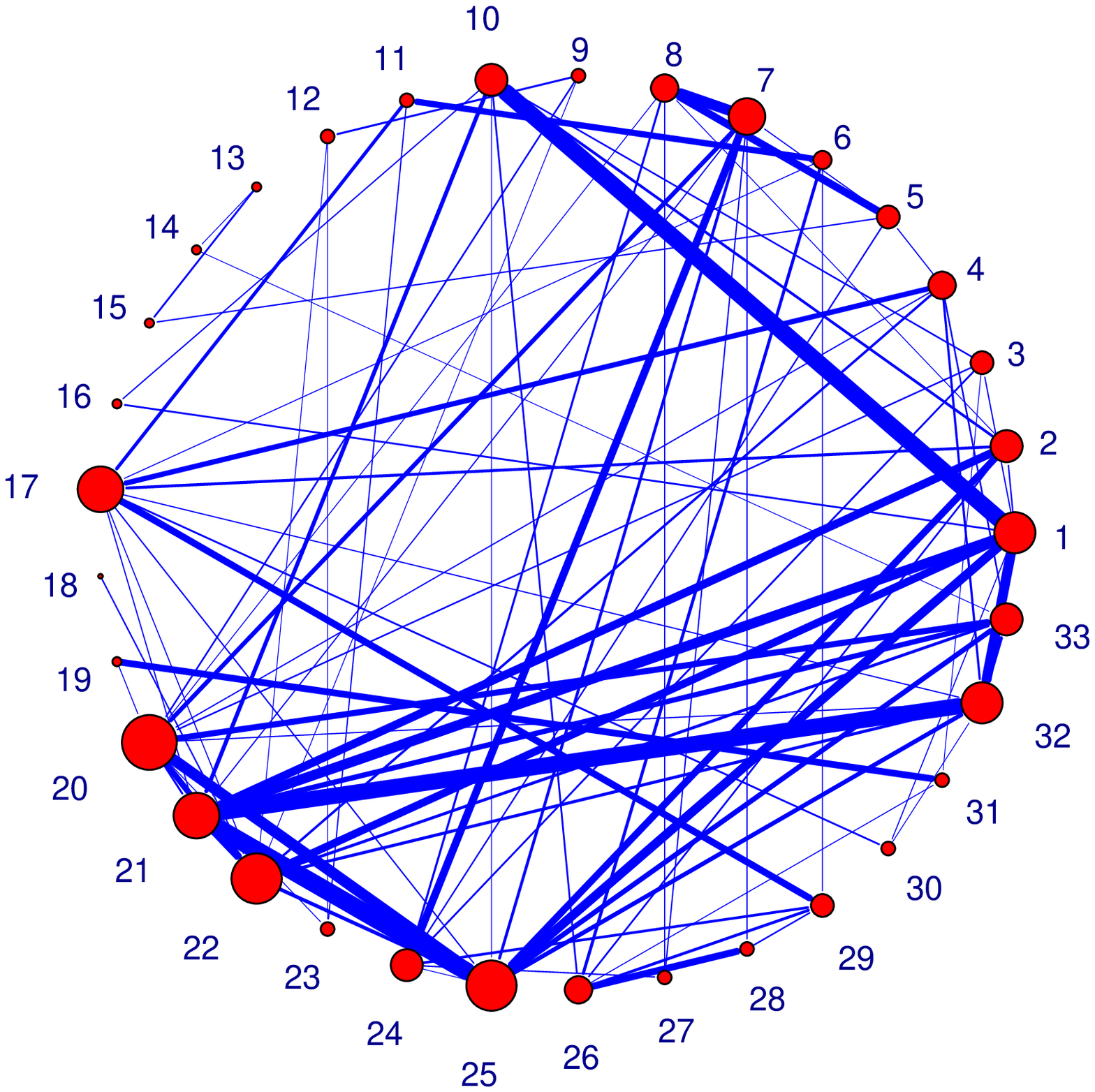} &
\includegraphics[width=0.45\textwidth]{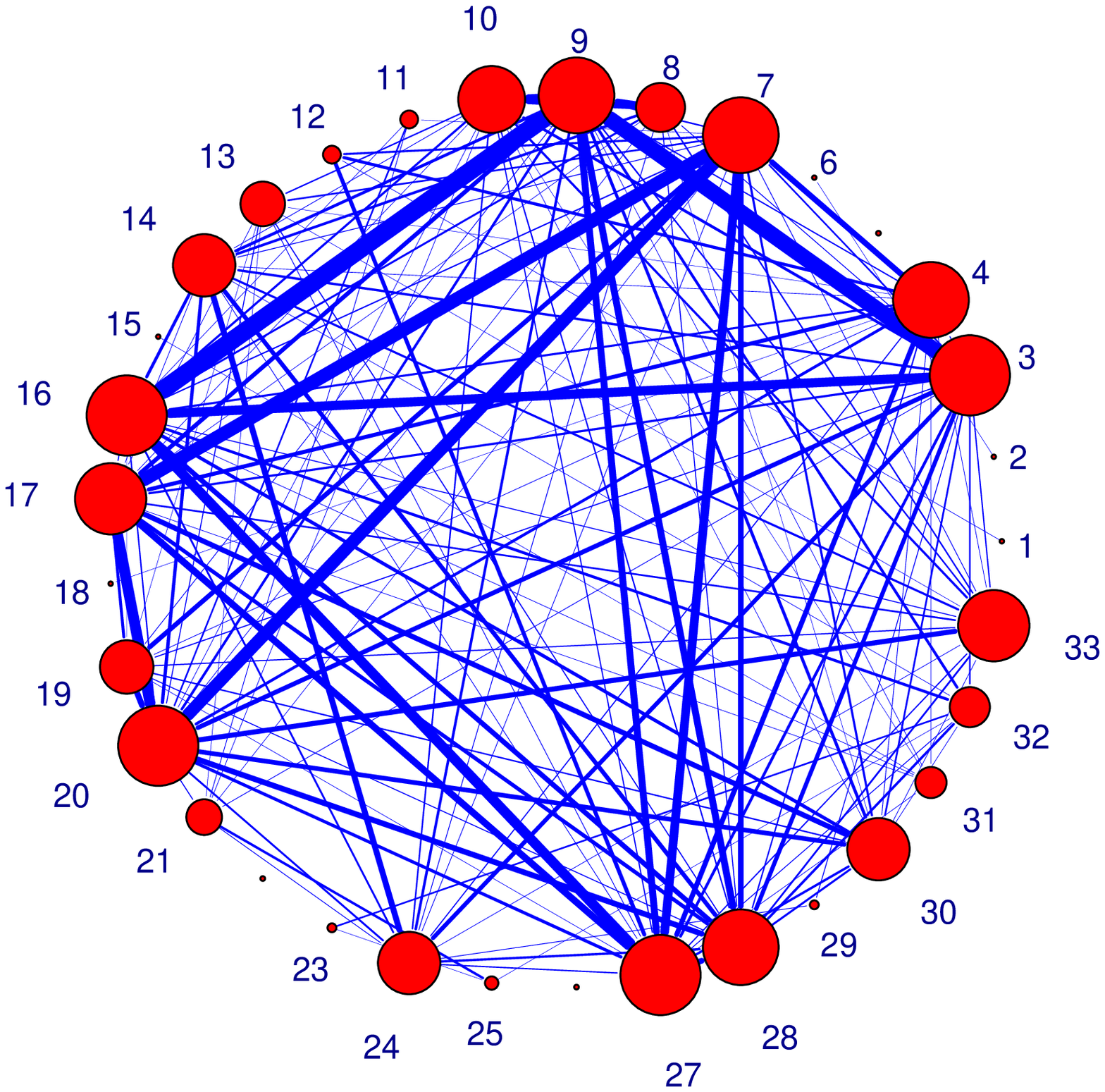} \\
(c) & (d) \\
\end{tabular}
\end{center}
\caption{Examples of topology of weighted coexpression networks: (a) GSE14333, Stage C (b) GSE14333, Stage D (c) GSE5206, Stage C (c) 17536/8, Stage C. Node numbering inherited from Tab.~\ref{tab:mapping}, node size and edge width proportional to degree and weight respectively.}
\label{fig:crcnets}
\end{figure}

In this last example, we want to assess the stability of a network inferred by high-throughput data in terms of distances between networks generated from data subsampling.
Sources of variability in this context are several: as a case study, here we consider three different publicly available (on GEO) microarray studies on the same pathology (colorectal cancer), on the same array platform (Affymetrix Human Genome U133 Plus 2.0 Array), with the same inference algorithm (WGCNA).
References and details on the three datasets are listed in Tab.~\ref{tab:crc}. The 33-genes signature from the paper \citep{smith10experimentally} (developed for differentiating Dukes' stage A and D and tested on stages B and C) are selected as the vertices of the subnetwork to infer.
\begin{table}[!ht]
\caption{Size of patient cohorts grouped by disease stage}
\label{tab:crc}
\begin{center}
\normalsize{
\begin{tabular}{c|ccc}
& \multicolumn{3}{c}{Reference and GEO Accession Number} \\
\cline{2-4}
&\citep{jorissen09metastasis} & \citep{kaiser07transcriptional} & \citep{smith10experimentally} \\
Dukes/AJCC Stage & GSE14333 & GSE5206 & GSE17536/GSE17538 \\
\hline
A/I&	44&	12&	28\\
B/II&	94&	32&	72\\
C/II&	91&	33&	76\\
D/IV&	61&	21&	56\\
\end{tabular}
}
\end{center}
\end{table}
The 33 genes map on 85 probes of the platform; during analysis, the expression of a gene is computed by averaging samplewise the expressions of all its mapping probes.
The list of the 33 genes included in the signature, together with the mapped probes, is included in Tab. \ref{tab:mapping}.
\begin{table}[!ht]
\caption{The 33-gene CRC signature described in \citep{smith10experimentally}, with the mapped probes on the Affymetrix Human Genome U133 Plus 2.0 Array platform.}
\label{tab:mapping}
\begin{center}
\tiny{
\begin{tabular}{r|l|l|l}
\# & Gene name & Ensemble ID & Mapped probes \\
\hline
1 &ACTB & ENSG00000075624 & 200801\_x\_at 213867\_x\_at 224594\_x\_at AFFX-HSAC07/X00351\_3\_at \\
&&& AFFX-HSAC07/X00351\_5\_at AFFX-HSAC07/X00351\_M\_at \\
2&DFNB31 &ENSG00000095397 & 221887\_s\_at 47553\_at \\
3&TMEM14A &ENSG00000096092 & 218477\_at \\
4&CIRBP &ENSG00000099622 & 200810\_s\_at 200811\_at 225191\_at 228519\_x\_at 230142\_s\_at \\
5&SYT17 &ENSG00000103528 & 205613\_at 229053\_at\\ 
6&AK1 &ENSG00000106992 & 202587\_s\_at 202588\_at \\
7&MGP &ENSG00000111341 & 202291\_s\_at 238481\_at \\
8&VDR &ENSG00000111424 & 204253\_s\_at 204254\_s\_at 204255\_s\_at 213692\_s\_at \\
9&C6orf64 &ENSG00000112167 & 218784\_s\_at 222741\_s\_at 232992\_at \\
10&HES1 &ENSG00000114315 & 203393\_at 203394\_s\_at 203395\_s\_at \\
11&TEX11 &ENSG00000120498 & 221259\_s\_at 233514\_x\_at 234296\_s\_at \\
12&MYOT &ENSG00000120729 & 219728\_at \\
13&EGR1 &ENSG00000120738 & 201693\_s\_at 201694\_s\_at 227404\_s\_at \\
14&DCTD &ENSG00000129187 & 201571\_s\_at 201572\_x\_at 210137\_s\_at \\
15&MMP13 &ENSG00000137745 & 205959\_at \\
16&TACC2 &ENSG00000138162 & 1570025\_at 1570546\_a\_at 202289\_s\_at 211382\_s\_at \\
17&CXCR7 &ENSG00000144476 & 1559114\_a\_at 212977\_at 232746\_at \\
18&DENND2A &ENSG00000146966 & 221885\_at 221886\_at 53991\_at \\
19&MUM1L1 &ENSG00000157502 & 229160\_at \\
20 &PDLIM5 &ENSG00000163110 & 203242\_s\_at 203243\_s\_at 211680\_at 211681\_s\_at 212412\_at \\
&&& 213684\_s\_at 216803\_at 216804\_s\_at 221994\_at 241208\_at \\
21&SPDYA &ENSG00000163806 & 238262\_at \\
22&NMNAT3 &ENSG00000163864 & 228090\_at 243738\_at \\
23&CRABP1 &ENSG00000166426 & 1563897\_at 205350\_at \\
24&ACYP2 &ENSG00000170634 & 206833\_s\_at \\
25&CSN3 &ENSG00000171209 & 207803\_s\_at \\
26&HPSE &ENSG00000173083 & 219403\_s\_at 222881\_at \\
27&STOX2 &ENSG00000173320 & 226822\_at 231969\_at 234317\_s\_at 234319\_at \\
28&SLC25A30 &ENSG00000174032 & 226782\_at 238171\_at \\
29&NQO1 &ENSG00000181019 & 201467\_s\_at 201468\_s\_at 210519\_s\_at \\
30&SPRY4 &ENSG00000187678 & 220983\_s\_at 221489\_s\_at \\
31&S100A3 &ENSG00000188015 & 206027\_at \\
32&PRTN3 &ENSG00000196415 & 207341\_at \\
33&HS3ST5 &ENSG00000249853 & 240479\_at \\
\end{tabular}
}
\end{center}
\end{table}
In Fig.~\ref{fig:crcnets} we show as an example three of the coexpression graphs (on the whole set of data) for stages C and D for three different datasets. The node numbering is taken from Tab.~\ref{tab:mapping}, the node size is proportional to its degree and the edge width is proportional to its weight.

To quantify network stability, for a given dataset and stage, we select a random fraction $p$ of the data and we generate the corresponding WGCNA; this procedure is repeated $N$ times, so to end up with $N$ WGCNA for each configuration (dataset, stage and $p$). 
Then all mutual $\binom{N}{2}=\frac{N(N-1)}{2}$ Ipsen-Mikhailov distances are computed, and for each set of $N$ graphs we build the corresponding distance histogram, reporting also mean and standard deviation. 
The lower the mean and the variance, the stabler the inferred network.

\begin{figure}[!ht]
\begin{center}
\begin{tabular}{cc}
\includegraphics[width=0.45\textwidth]{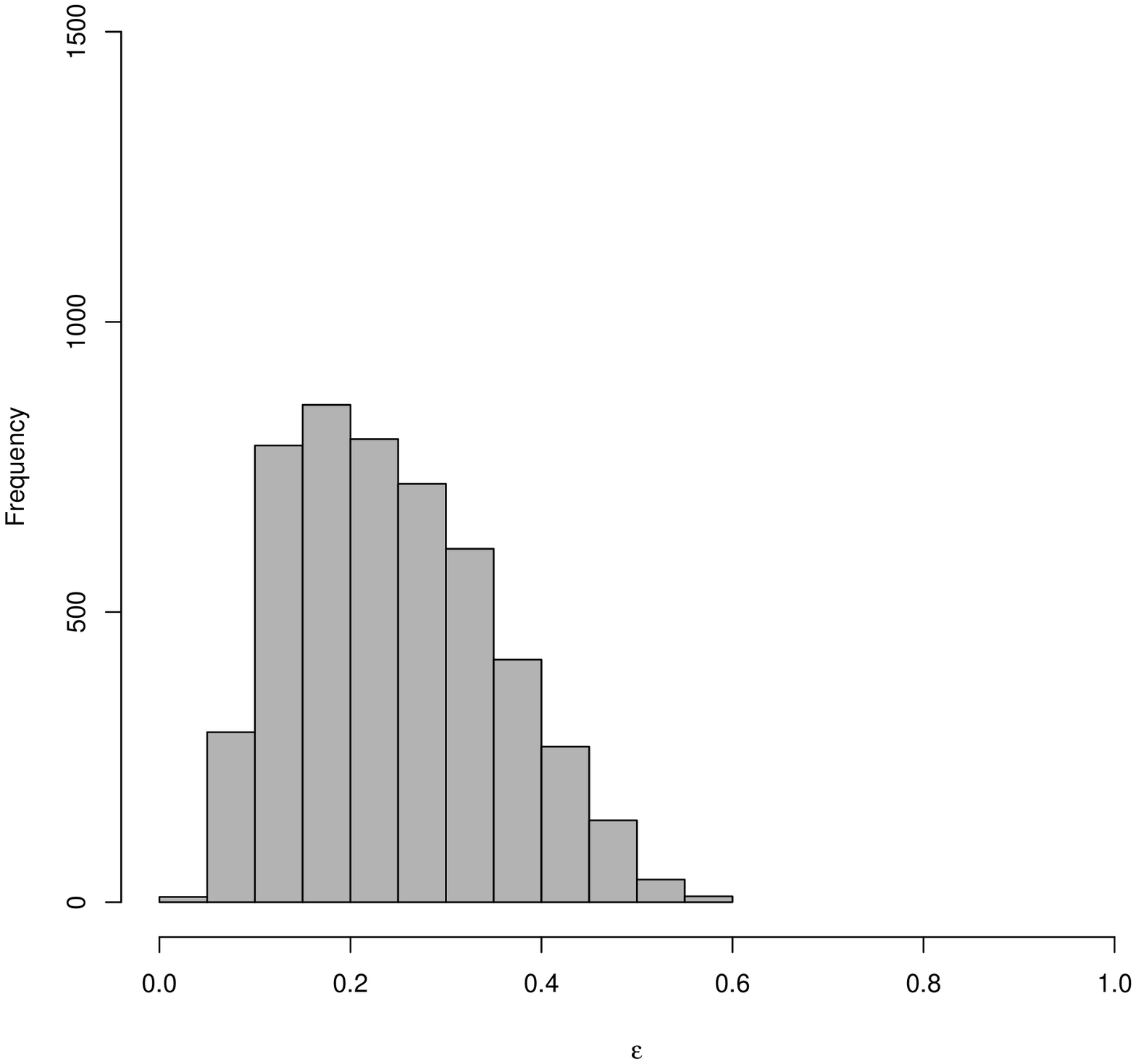} 	&
\includegraphics[width=0.45\textwidth]{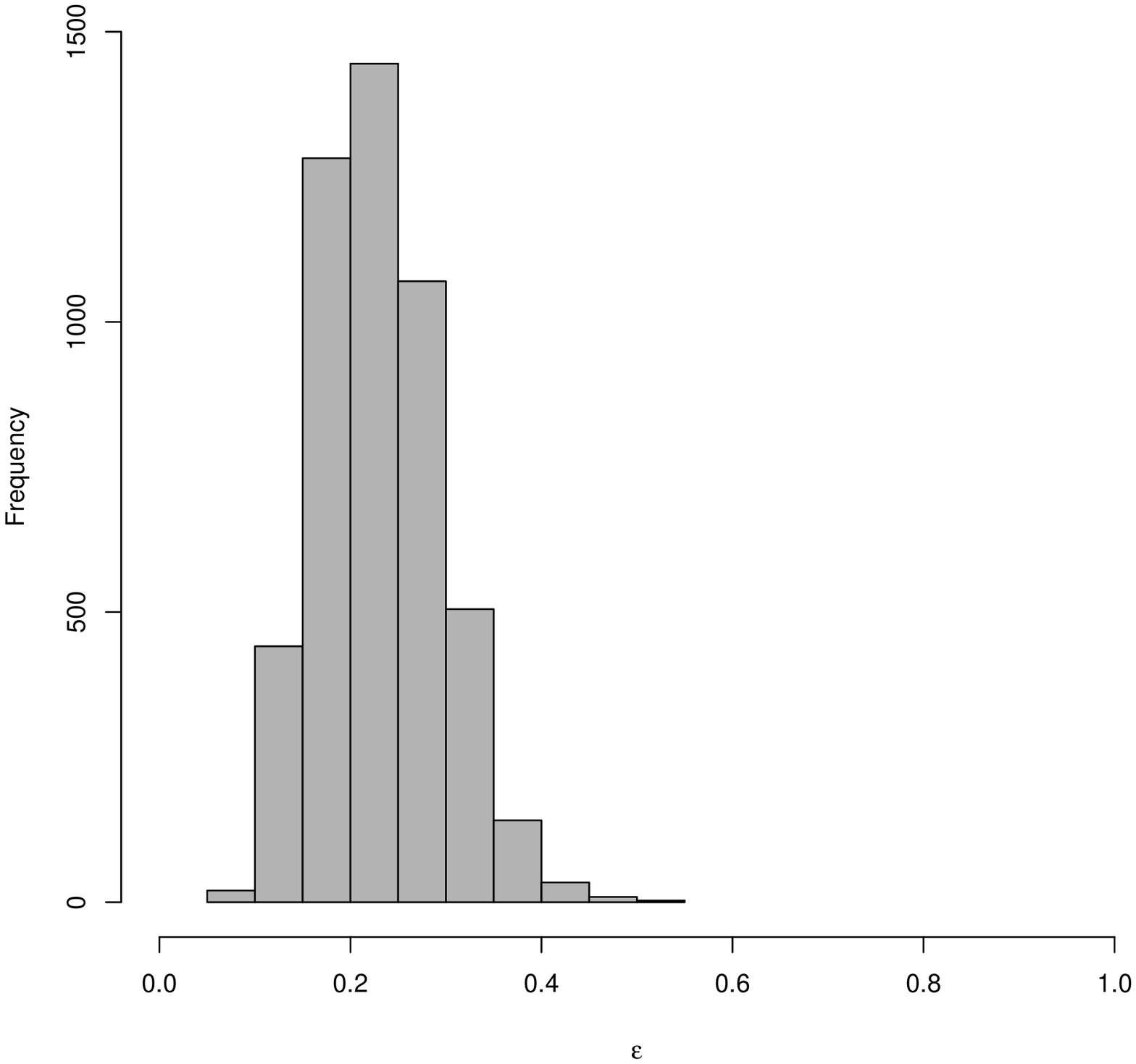} 	\\
(a) & (b) \\
\includegraphics[width=0.45\textwidth]{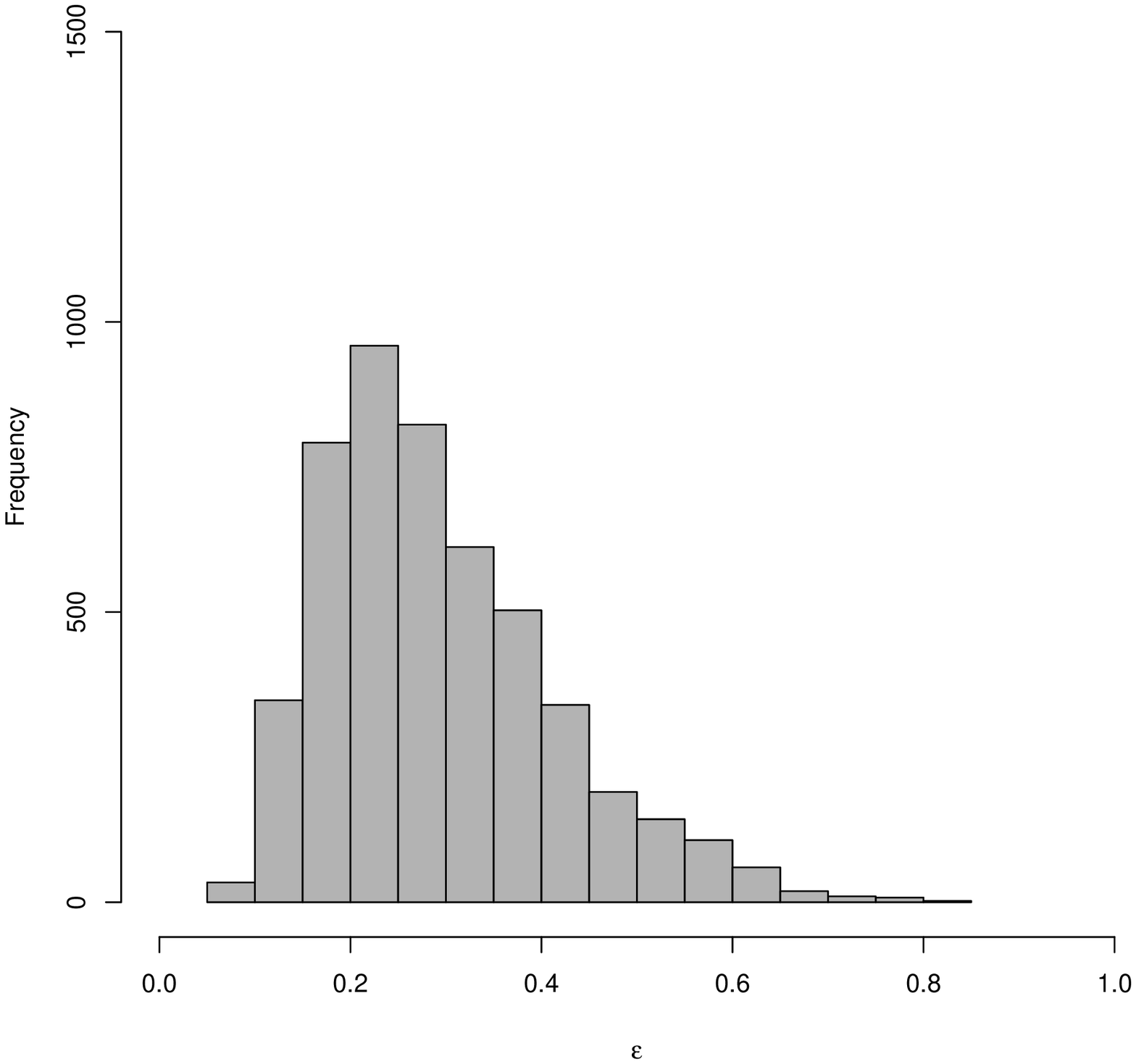} 	&
\includegraphics[width=0.45\textwidth]{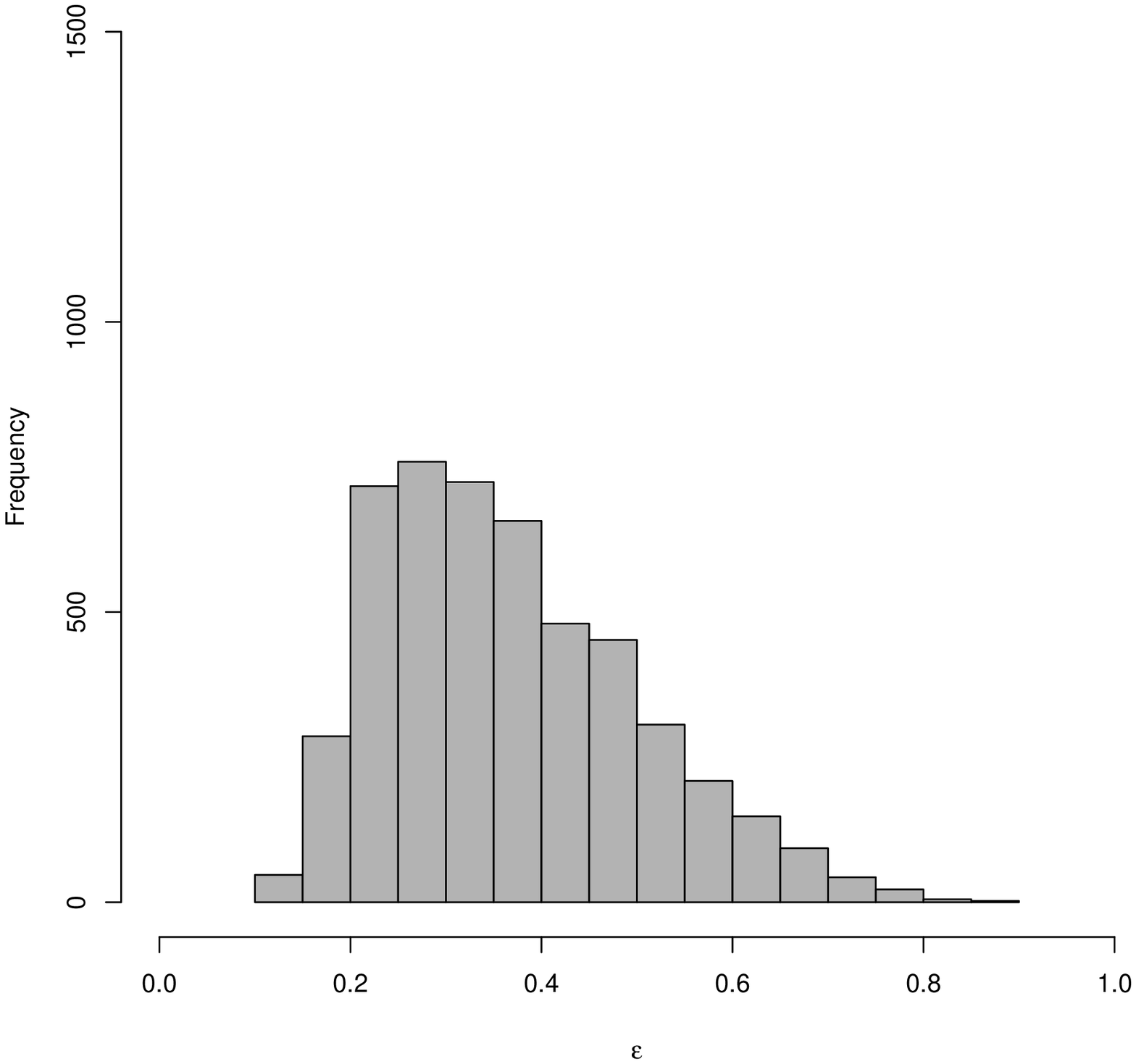} 	\\
(c) & (d) \\
\end{tabular}
\end{center}
\caption{Histogram of distance distributions for four of the cases listed in Tab.~\ref{tab:means}: (a) GSE14333, Stage D, $p=\frac{2}{3}$, (b) GSE17536/8, Stage B, $p=\frac{2}{3}$, (c) GSE5206, Stage C, $p=\frac{2}{3}$, (d) GSE17536/8, Stage A, $p=\frac{1}{2}$.}
\label{fig:hists}
\end{figure}

In Tab.~\ref{tab:means} we show the results for $p=\frac{1}{2}$ and $p=\frac{2}{3}$, with $N=100$ replicates.
\begin{table}[ht]
\caption{Mean and variance for the sets of 4950 distances for all combinations of dataset and stage, in the two cases of using 2/3 and 1/2 of the data.}
\label{tab:means}
\begin{center}
\begin{tabular}{l|cc|cc|cc|cc}
$p=\frac{2}{3}$ & 
\multicolumn{2}{c|}{Stage A} & \multicolumn{2}{c|}{Stage B} & \multicolumn{2}{c|}{Stage C} & \multicolumn{2}{c}{Stage D} \\
 & $\mu$ & $\sigma^2$ & $\mu$ & $\sigma^2$ & $\mu$ & $\sigma^2$ & $\mu$ & $\sigma^2$ \\
\hline
GSE14333 & 0.1988 & 0.0065 & 0.1787 & 0.0088 & 0.1843 & 0.0086 & 0.2447 & 0.0112 \\
GSE5206 & 0.2457 & 0.0073 & 0.2180 & 0.0051 & 0.2777 & 0.0137 & 0.2409 & 0.0086 \\
GSE17536/8 & 0.3306 & 0.0169 & 0.2301 & 0.0040 & 0.2199 & 0.0028 & 0.2173 & 0.0038 \\
\hline
\hline
$p=\frac{1}{2}$ & 
\multicolumn{2}{c|}{Stage A} & \multicolumn{2}{c|}{Stage B} & \multicolumn{2}{c|}{Stage C} & \multicolumn{2}{c}{Stage D} \\
 & $\mu$ & $\sigma^2$ & $\mu$ & $\sigma^2$ & $\mu$ & $\sigma^2$ & $\mu$ & $\sigma^2$ \\
\hline
GSE14333 & 0.2075 & 0.0066 & 0.2135 & 0.0112 & 0.2028 & 0.0104 & 0.2978 & 0.0159 \\
GSE5206 & 0.2602 & 0.0070 & 0.2464 & 0.0071 & 0.2931 & 0.0146 & 0.2596 & 0.0095 \\
GSE17536/8 & 0.3668 & 0.0176 & 0.2681 & 0.0048 & 0.2646 & 0.0044 & 0.2483 & 0.0061 \\
\hline
\end{tabular}
\end{center}
\end{table}
For stages A, B and C the best stability is detected on the GSE14333 dataset, while for stage D GSE175536/8 results the stabler dataset. 
Moreover, the $(\mu,\sigma^2)$ couples listed in Tab.~\ref{tab:means} do not show a great variability among the 24 listed cases.
A larger range of situations can be appreciated by looking at the shapes of the distribution of each set of 4950 distances. 
In fact, although some of the cases are almost gaussian-like, a number of other combinations of dataset and stage are represented by very skewed distribution: for them, considering mean and variance as descriptive parameters may be interpretatively misleading.
As an example, GSE14333, Stage D and GSE17536/8, Stage B for $p=\frac{2}{3}$ have rather similar mean and variance, but a quite different distribution shape as shown by Fig.~\ref{fig:hists}).
We can conclude observing that the plotted histograms show how in several cases the infered network can be heavily dependent on the particular chosen subset of data, leaving the network built on the whole dataset affected by a relatively large level of uncertainty (instability): this may be due both to high variability in the data distribution, but also to high sensibility of the algorithm to data perturbation.
This fact should always be taken into account when assessing the reliability of a reconstructed network in order to avoid drawing biological consideration from a possible false positive edge linking two nodes.

\section{DISCUSSION}
\label{sec:discussion}
Ipsen-Mikhailov $\epsilon$ distance is an effective metric for comparing (biological) networks in various situations. 
Its definition involves the distribution of the Laplacian spectrum of the networks, so it deal with the structure of the underlying graph, rather than focussing on the local pattern of the wiring differences.
It is mostly consistent with more classification measures such as MCC, but it allows detection of finer differences.
The presented examples show effectiveness and usefulness of $\epsilon$ in different biological tasks, but additional applications can be considered wherever a quantitative network comparison is needed.
Finally, the use of Ipsen-Mikhailov distance in Transcription Starting Sites network will be presented within the Fantom5\footnote{\url{http://www.riken.go.jp/engn/r-world/info/info/2010/101206/index.html}} initiative led by Riken Institute.

\section*{ACKNOWLEDGEMENTS}
The authors acknowledge funding by the European Union FP7 Project HiperDART and by the PAT funded Project ENVIROCHANGE.

\section*{DISCLOSURE STATEMENT}
No competing financial interests exist.

\bibliography{jurman11ipsen}
\end{document}